\documentclass[aps,onecolumn,preprintnumbers,superscriptaddress]{revtex4}
\usepackage{graphicx} % Required for inserting images
\usepackage{amsmath}
\usepackage{amsfonts}
\usepackage{amssymb}
\usepackage{physics}
\usepackage{hyperref}
\usepackage{bm}
\usepackage{xcolor}
\usepackage{subfigure}
\usepackage{tikz}
\usepackage{siunitx}
\usepackage{pgffor}
\usepackage{pgfplots}
\usetikzlibrary{positioning}
\usetikzlibrary{arrows.meta}
\usetikzlibrary{arrows}
\usetikzlibrary{calc}
\usetikzlibrary{fit}
\usepackage{tikz-3dplot}
\usetikzlibrary{3d, calc}

\def\be{\begin{equation}}
\def\ee{\end{equation}}
\def\bea{\begin{eqnarray}}
\def\eea{\end{eqnarray}}

\begin{document}
\author{Shane Thompson}
\email{shane.n.thompson.ctr@us.navy.mil}
\affiliation{NRC Research Associate, U.S. Naval Research Laboratory}
\author{Daniel Gunlycke}
\email{lennart.d.gunlycke.civ@us.navy.mil}
\affiliation{U.S. Naval Research Laboratory}
% \author{Shane Thompson}
% \email{shane.n.thompson.ctr@us.navy.mil}
% \affiliation{U.S. Naval Research Laboratory}
% \author{Daniel Gunlycke}
% \email{lennart.d.gunlycke.civ@us.navy.mil}
% \affiliation{U.S. Naval Research Laboratory}

\begin{abstract}
    Accurate ground-state energy calculations remain a central challenge in quantum 
chemistry due to the exponential scaling of the many-body Hilbert space. Variational 
Monte Carlo and variational quantum eigensolvers offer promising ansatz optimization 
approaches but face limitations in convergence as well as hardware constraints. We 
introduce a particular Selected Configuration Interaction (SCI) algorithm that uses auto-regressive neural networks (ARNNs) to guide subspace expansion for ground-state 
search. Leveraging the unique properties of ARNNs, our algorithm efficiently constructs 
compact variational subspaces from learned ground-state statistics, which in turn 
accelerates convergence to the ground-state energy. Benchmarks on molecular 
systems demonstrate that ARNN-guided subspace expansion combines the strengths 
of neural-network representations and classical subspace methods, providing a scalable 
framework for classical and hybrid quantum-classical algorithms.
\end{abstract}
\title{Auto-regressive Neural Quantum State Sampling for Selected Configuration Interaction}

\maketitle
\section{Introduction}

The accurate computation of ground-state energies of molecular systems is a fundamental goal in quantum chemistry, as it serves as a foundation in predicting molecular stability, thermodynamics, and reactivity. Classical approaches such as configuration interaction (CI) and the coupled cluster (CC) method \cite{cizek1966} have demonstrated remarkable success in realizing this goal, especially in the case of weakly correlated systems. However, the steep computational scaling of full configuration interaction (FCI) with system size—combinatorial in the number of spin-orbitals and electrons \cite{Szabo1996}—limits its applicability to small molecules or reduced basis sets. While the scaling is not as severe for coupled cluster, its extension to multi-reference systems is limited \cite{McArdle2020}. As a result, alternative paradigms for representing and optimizing many-body wavefunctions have become increasingly important.

Several other classical methods offer partial solutions to this scaling problem. Density functional theory (DFT) \cite{Kohn1965} remains the most widely used method due to its relatively low computational cost and ability to handle large systems. However, the lack of systematic improvability and poor performance in strongly correlated regimes limit its utility for high-accuracy applications \cite{Jones2015}. Density matrix renormalization group (DMRG) \cite{White1992,Chan2011} methods have become the method of choice for quasi-one-dimensional systems and systems with low entanglement, offering near-exact ground states for strongly-correlated Hamiltonians. Yet DMRG's efficiency decreases in more entangled or higher-dimensional problems, motivating the continued search for scalable and accurate alternatives.

Two additional variational approaches have emerged in recent years. On the one hand, quantum simulation techniques such as the variational quantum eigensolver (VQE) algorithm \cite{Peruzzo_2014,Tilly2022} have attracted attention for their potential to solve quantum chemistry problems on near-term quantum devices. VQE uses a parameterized quantum circuit to prepare trial wavefunctions and evaluates energies via quantum measurement. In principle, quantum hardware can efficiently represent states in exponentially large Hilbert spaces while circumventing the need for an approximate theory. However, in practice, VQE operates within a truncated variational manifold defined by the structure and depth of the quantum-circuit ansatz. This ansatz-dependent restriction is often necessary due to hardware noise and limited coherence times. Smaller ans\"atze like the Hardware-Efficient ansatz \cite{Kandala_2017} occupy an exponentially small and physically irrelevant corner of the full Hilbert space \cite{McArdle2020} while, even in the absence of machine noise, a deeper quantum circuit can quite easily suffer from a severe barren plateau problem \cite{McClean2018,Larocca2025}.

% Even in the case of highly expressive and accurate ans\"atze, one also encounters the computational cost of evaluating expectation values over complex Hamiltonians with many terms. This considerable measurement overhead typically calls for grouping techniques which attempt to reduce the number of circuits to be processed while still respecting statistical variance  \cite{Tilly2022}. 

On the other hand, variational Monte Carlo (VMC) methods, particularly those incorporating Neural Quantum States (NQS), offer a fully classical route to variational energy minimization with a flexible ansatz \cite{Carleo2017,Saito2017,Pfau2020}. Neural networks have the ability to efficiently approximate quantum states by encoding its complexity, over both amplitudes and phases, into a relatively small set of weights defining the variational manifold. Furthermore, auto-regressive architectures \cite{Uria2016,germain2015,vaswani2023} in particular enable direct, unbiased sampling from the probability distribution given by the square of the wavefunction \cite{Sharir2020}, i.e. the ``Born probabilities" induced by the Born rule. VMC is scalable to relatively large systems and does not suffer from the severe sign problem that appears in projective Monte Carlo frameworks \cite{Troyer2005,Booth2009}. This is due to the ability to place all phase information in a ``local estimator" whose expectation value is over a traditional probability distribution determined by just the amplitudes of the NQS. Details on the VMC approach are briefly discussed in Appendix \ref{appendix:VMC}.

VMC does come with its own set of shortcomings. As might be expected, its accuracy is determined by the expressive power of the ansatz, and in strongly-correlated systems, accurate representations may require large parameter counts and deep architectures making the training process more challenging, requiring properly handled initialization of the network parameters. Another issue is that expectation values and their gradients are estimated through sampling performed in the computational basis, which could lead to additional convergence issues associated with variance in the local estimator over the exponentially large space \cite{Umrigar1988,Trail2008,Sinibaldi2023}. 

In the spirit of achieving an efficient representation of a quantum state, a set of subspace expansion techniques within the class of methods known as ``selected configuration interaction" (SCI), itself developed outside the context of NQS \cite{Huron1973,Tubman2016,Holmes2016}, have proven remarkably effective. These methods construct a compact variational subspace by iteratively selecting basis states of electronic configurations that contribute significantly to lowering the energy, typically based on perturbative criteria or matrix element heuristics. Although quite successful, these selection rules can still miss important configurations and subtle correlation effects.

In this work, we propose the use of auto-regressive neural networks (ARNNs) to guide subspace expansion in quantum chemistry. Rather than relying on Hamiltonian-induced transitions to guide the subspace search, we use an ARNN to model the current-iteration approximation to the ground-state Born probabilities and evaluate the probability assigned by the model to candidate electronic configurations. The configurations need not appear in the data samples obtained from the approximation the network was trained on, nor even in the approximation itself. ARNNs provide exact and normalized probability distributions over discrete configuration spaces, and naturally incorporates temperature scaling, a useful sampling technique previously seen in some classification tasks \cite{hinton2015,Guo2017}. Temperature scaling could also be used as an importance sampling strategy in VMC \cite{misery2025}. This technique allows the sampling distribution to be sharpened or broadened in a controlled manner and thus enables targeted exploration of configurations with non-negligible amplitude in the subspace-expansion process, significantly improving the efficiency of subspace selection. 

Similar machine learning approaches to SCI exist \cite{Coe2018,Herzog2023,Bilous2025}, yet our work still contains important differences. Ref. \cite{Coe2018} uses regression to estimate updated configuration coefficients, but must evaluate them all explicitly for configuration ``pruning." Ref. \cite{Herzog2023} is much closer to our work and employs a Restricted Boltzmann Machine (RBM) as a generative model to prepare samples of candidate configurations. The most important distinctions that one can draw between this work from ours are that (1) Ref. \cite{Herzog2023} relies on Gibbs sampling instead of the fast, unbiased, and locally-guided sampling promised by auto-regressive architectures \cite{Barrett2022}, (2) the authors rule out any possible benefits of temperature scaling, yet in this paper we demonstrate its strong potential for configuration discovery. Most of all, (3) the authors initialize the procedure with a Configuration Interaction Singles and Doubles (CISD) ground-state approximation which, as demonstrated in our results, limits the capabilities of the method.

In Ref. \cite{Bilous2025}, a binary classifier was utilized to iteratively select ``important" and ``unimportant" configurations from pools of proposed configurations. However, instead of separately identifying and then classifying proposal states, we use the distribution data itself to find and score new basis states for inclusion in the variational subspace, in line with recent stochastic SCI techniques \cite{kanno2023,Gunlycke2024,gunlycke2025,Moreno2025}. By integrating ARNNs into the subspace selection loop, we aim to combine the expressivity and scalability of neural-network states with the systematic convergence tendencies of SCI-like expansions. We benchmark this algorithm on a set of representative molecular systems, and show that ARNN-guided subspace expansion accelerates convergence to chemically accurate ground-state energies and provides a flexible framework for incorporating machine-learned insight into classical and hybrid quantum-classical algorithms.

In Sec. \ref{section:Description_of_Algorithm}, we outline our iterative procedure and follow up with some scaling considerations. We describe the specifics for each step of the algorithm and focus the discussion toward quantum chemistry applications, although the algorithm can be applied more generally. In Sec. \ref{section:Results} we show results for some small-scale molecular systems (between $24$ and $36$ spin-orbitals) and demonstrate the effective performance of the algorithm. We conclude in Sec. \ref{section:Conclusion}.

\section{Description of Algorithm} \label{section:Description_of_Algorithm}
The procedure proceeds according to the following steps:
\begin{enumerate}
    \item Construct an approximation $\ket{\Psi_\text{init}}$ of the molecular ground state (GS) from which we can sample configurations $n=(n_0,n_1,\ldots,n_{M-1})$, in the computational basis of electronic configurations. The approximation could have a classical representation or even be a prepared quantum state on quantum hardware.
    \item Sample this state in the computational basis to generate training data.
    \item Train a neural network against this data, using a number of training samples $N_\text{T}$. This provides a rough NQS approximation of the quantum state that was sampled. Since the data is only in the computational basis, the phases appearing in the quantum state are arbitrary.
    \item Sample the neural network, again in the computational basis and with a set number of network samples $N_\text{N}$, and collect a specified number $N_\text{U}$ of unique configurations.
    \item Diagonalize the $N_\text{U}\times N_\text{U}$ Hamiltonian in the subspace of unique configurations and determine its lowest-energy state to obtain a new sparse approximation of the true GS.
    \item Repeat Steps $1-5$ for each iteration $i$, replacing $\ket{\Psi_\text{init}}$ with the new approximation $\ket{\Psi_{i-1}}$ from Step 5 in iteration $i-1$, until convergence criteria are satisfied and we accept the final state $\ket{\Psi_\text{opt}}$.
\end{enumerate}

\begin{figure}[t]
\begin{center}
\includegraphics[]{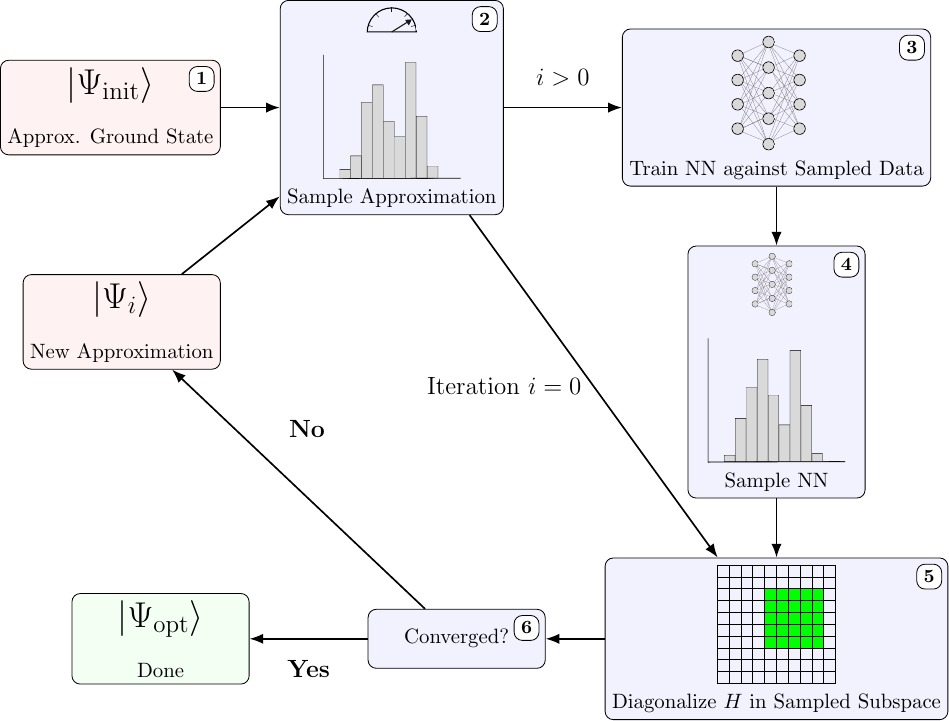}
\caption{NQS-based Selected Configuration Interaction procedure for approximating molecular ground-state energies and wavefunctions. An initial GS approximation $\ket{\Psi_\text{init}}$ generates training data (configurations with sampling frequencies) from which a neural network can be trained. Sampling the neural network adds unseen configurations to a ``sampled subspace" over which we perform exact diagonalization and obtain a new GS approximation, which we either accept as $\ket{\Psi_\text{opt}}$ or use as an initiator $\ket{\Psi_i}$ for the next iteration. For $i=0$, we skip Steps 3 and 4, letting the approximation itself determine the sampled subspace.}
\label{fig:flowchart}
\end{center}
\end{figure}

\begin{figure}[t]
    \centering
    \includegraphics[width=0.45\linewidth]{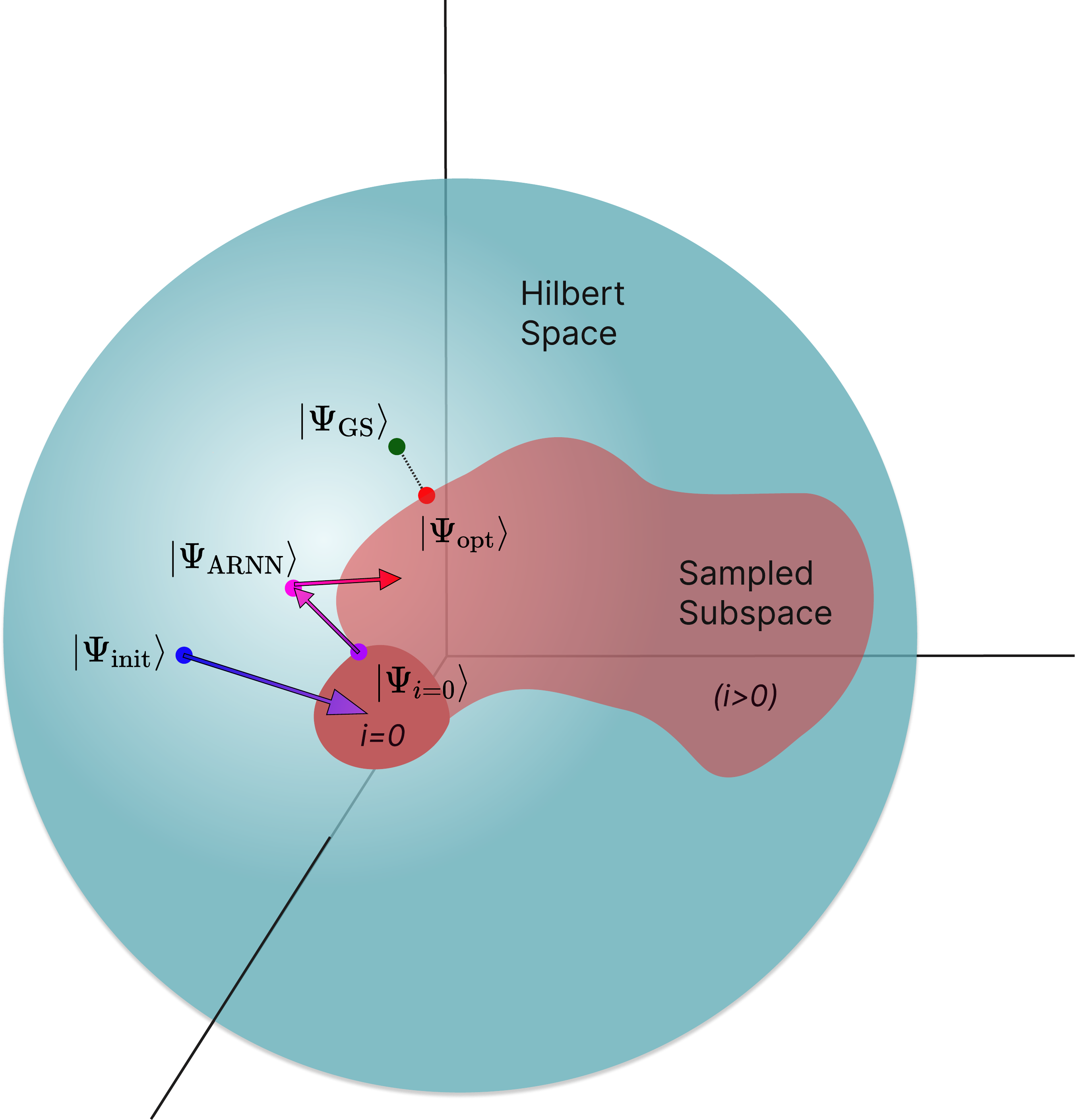}
    \caption{Visual representation of iterations $i=0$ and $i=1$ of our algorithm. The initial state $\ket{\Psi_{\text{init}}}$ is sampled for important configurations and an efficient approximation $\ket{\Psi_{i=0}}$ is constructed by diagonalizing the Hamiltonian in the sampled subspace. The NQS $\ket{\Psi_{\text{ARNN}}}$ is trained from $\ket{\Psi_{i=0}}$ and then is itself sampled. $\ket{\Psi_{\text{opt}}}$ is an energy minimum in the sampled subspace and therefore approximates the true ground state, $\ket{\Psi_{\text{GS}}}$. The three arrows correspond to (1) obtaining the $i=0$ subspace from $\ket{\Psi_\text{init}}$, (2) training $\ket{\Psi_\text{ARNN}}$ using data from $\ket{\Psi_{i=0}}$, and (3) obtaining the $i>0$ subspace from $\ket{\Psi_\text{ARNN}}$.}
    \label{fig:NQS_diagram}
\end{figure}

The algorithm is illustrated in Fig. \ref{fig:flowchart}. A visual representation of the first iteration is shown in Fig. \ref{fig:NQS_diagram}. Note that in the latter, there is no requirement for the NQS $\ket{\Psi_{\text{ARNN}}}$ to lie in a physical subspace associated with known GS symmetries; we can always discard any samples corresponding to unphysical states. We also note that the prescription above suggests that computational-basis measurements from the initial approximation carry over as raw data for training the network. In practice, however, it turns out that the inclusion of an ``iteration zero", in which we immediately perform subspace diagonalization using the raw sampling data (without using the network), yields a more effective version of the algorithm which is used to generate the results in Sec. \ref{section:Results}. In Fig. \ref{fig:flowchart}, iteration zero is indicated by an arrow for $i=0$ which bypasses the steps involving the NN and limits the subspace to the configurations discovered directly from $\ket{\Psi_\text{init}}$. Iteration zero condenses the original sampled data in a way that assists later iterations, when the network is actually used.

Before we go into the details of each step, we describe how this procedure scales reasonably well with system size. In Step 1, we request only that we pick a state which can be efficiently sampled enough times in Step 2 to obtain a rough sketch of its probability distribution in the computational basis. It could be a state that is sparse in this basis (e.g. the Hartree--Fock configuration), a matrix product state (MPS) approximation of the GS, which can be sampled efficiently, or a quantum circuit approximation obtained from some quantum simulation, say the VQE algorithm. In Ref. \cite{Bennewitz2022}, VQE was used to initialize network parameters for Variational Monte Carlo, and indeed the results in Sec. \ref{section:Results} suggest that wavefunction approximations more true to the nature of the ground state are advantageous for our algorithm as well, in certain regimes.

The trained NQS in Step 3 would benefit greatly from large amounts of measurement data prepared in Step 2. But this is not practical in terms of both sample preparation and network training. Instead, we control the scaling of the amount of data we use, and then rely on one of the key advantages of neural networks -- their generalization power. In our results, we also restrict the number of training samples to some integer $N_\text{T}$, which can be kept quite low thanks to our use of small sampled subspaces as well as a technique known as temperature scaling, to be discussed in Section \ref{Section:Sampling}. One could additionally assume that the data contains noise due to non-ideal sampling conditions, but in this paper we restrict the discussion to the ideal case.

In Step 4, we have the ability to efficiently sample our new classical neural network, extracting important information about $\ket{\Psi_\text{init}}$, $\ket{\Psi_i}$, or even the exact GS. While we have far more sampling freedom as opposed to, for example, performing shots on limited quantum hardware, we still desire that the number of required samples $N_\text{N}$ does not scale uncontrollably with system size. Noting that some systems, particularly in quantum chemistry, have the potential to exhibit a degree of sparsity in the computational basis \cite{Ivanic2001,Anderson2018}, one can limit the number of necessary shots according to this sparsity, and even more so using the sampling flexibility associated with neural networks.

We also need to make sure that the number of unique configurations $N_\text{U}$ we keep from the NN samples remains low enough so that Hamiltonian diagonalization in the subspace formed by these states (Step 5) remains tractable. For the molecular systems considered in Sec. \ref{section:Results}, we track the ratio of subspace dimension to the dimension of the full Hilbert space, as we want to ensure that this number quickly vanishes for larger systems. We also note that once we have chosen a subspace formed from $N_\text{U}$ unique configurations, it is typically possible to construct the $N_\text{U}\times N_\text{U}$ Hamiltonian in this subspace given that e.g. quantum chemistry Hamiltonians contain a number of terms that is polynomial in the size of the basis set used. 

Our algorithm remains iterative as long as the output state from Step 5 can be efficiently sampled, just like the initial approximation $\ket{\Psi_{\text{init}}}$, for all iterations. This is guaranteed as long as the ground-state energy can be estimated to within chemical accuracy using a sparse wavefunction representation in the computational basis (which, as noted, may be possible). This is our primary assumption in this work, and this feature can be exploited, as shown below.

\subsection{Second-Quantized Hamiltonian and Orbital Mapping}

In quantum chemistry, molecular systems are typically described by electronic Hamiltonians in second quantization, expressed in terms of fermionic creation and annihilation operators. This formulation provides a natural framework for working with discrete Fock states and is widely used in many classical methods including Configuration Interaction (CI), exact diagonalization (ED), coupled cluster (CC), and the density matrix renormalization group (DMRG) algorithm. It is also naturally suited for quantum algorithms like the Variational Quantum Eigensolver (VQE) and Cascaded Variational Quantum Eigensolver (CVQE) \cite{Gunlycke2024} algorithms, and has recently been adopted in Neural Quantum State (NQS) techniques \cite{Choo2020,Barrett2022}.

The first step is to choose a finite single-particle basis set, which defines the space in which electronic wavefunctions are expanded. These basis functions are typically atom-centered and can range from minimal sets such as STO-3G \cite{Hehre1969} to more expressive families like cc-pVDZ or aug-cc-pVTZ \cite{Dunning1989,Kendall1992}. Once a set is chosen, one constructs molecular orbitals—usually as linear combinations of basis functions by solving the Hartree-Fock (HF) equations.

Using the HF molecular orbitals as a working basis, the electronic Hamiltonian (in the Born-Oppenheimer approximation) takes the general second-quantized form \cite{McArdle2020}:
\be\label{eq:BO_ham}
H = \sum_{pq} h_{pq} \, a_p^\dagger a_q + \frac{1}{2} \sum_{pqrs} h_{pqrs} \, a_p^\dagger a_q^\dagger a_r a_s,
\ee
where \(a_q^\dagger\) and \(a_q\) are fermionic creation and annihilation operators associated with molecular ``spin-orbital" $q$ satisfying the canonical anti-commutation relations $\{a_p,a_q^\dagger\} = \delta_{p,q},\ \{a_p,a_q\} = 0$, with $\{A,B\}\equiv AB+BA$. Here, the term ``spin-orbital" is defined in Refs. \cite{Szabo1996,McArdle2020} as the orthogonal single-electron wavefunctions available for occupation by the electrons in the system (like sites on a lattice). The electronic configurations we look for are Fock (basis) states $n\leftrightarrow\ket{n_0,n_1,\ldots,n_{M-1}}$ formed by applying $N_\text{e}$ $a_q^\dagger$'s to the fermionic vacuum, where $M$ and $N_\text{e}$ are the number of spin-orbitals and electrons, respectively. The many-electron wavefunctions for these Fock states are given by Slater determinants over the occupied single-electron wavefunctions. 

In Eq. \eqref{eq:BO_ham}, \(h_{pq}\) and \(h_{pqrs}\) are the one- and two-electron molecular integrals, respectively. These integrals are computed efficiently based on the chosen basis set and nuclear geometry, and they encode all electron-nuclear and electron-electron interactions. The problem of finding the ground state of fermionic Hamiltonians of the form \eqref{eq:BO_ham} is QMA-complete in the worst case, which would preclude even a solution using a quantum computer, yet many problems in quantum chemistry may still fall within regimes in which efficient simulation, using classical and/or quantum resources, remains possible \cite{Whitfield2013,McArdle2020}.

The Hartree--Fock procedure yields a single-determinant approximation to the ground state and serves as a natural reference point for more sophisticated treatments. While multi-configurational self-consistent field (MCSCF) methods like Complete Active Space Self-Consistent Field (CASSCF) \cite{Roos1980,Schmidt1998} provide systematic improvements over HF by optimizing orbital wavefunction descriptions in a way that is consistent with strong static correlations, we do not employ them in this work. Instead, we use the HF reference as a starting point for constructing the Fock basis and  Hamiltonian used in our procedure. Furthermore, our results are restricted to smaller basis sets, namely STO-3G and 6-31g, to limit computational cost and focus on proof-of-concept demonstrations.

As is often done in NQS approximations of quantum states \cite{Choo2020,Barrett2022}, we use configurations $n$, represented by bitstrings, as the inputs for a neural network. Each input corresponds to one of the $M$ molecular spin-orbitals, and the binary input vector $(n_0,n_1,\ldots,n_{M-1})$ encodes the occupation (0 or 1) of each spin-orbital in a given configuration. We adopt an orbital-ordering convention similar to the block-ordered Jordan-Wigner transformation used in IBM's \textit{Qiskit Nature} \cite{qiskit_nature}. When reading the binary input vector, the first half of the inputs correspond to spin-down orbitals, and the second half correspond to spin-up orbitals. Configurations read-out left-to-right in binary-vector notation correspond to a top-to-bottom mapping on the network (see Fig. \ref{fig:our_ARNN}). Therefore, assigning neuron $0$ to be at the top of the network, neurons $0$ to $M/2-1$ correspond to spin-down molecular orbitals of decreasing energy, while neurons $M/2$ to $M-1$ correspond to spin-up molecular orbitals of decreasing energy. For example, in a Hartree--Fock molecular-orbital basis the HF reference state itself (in the spin-balanced case) can be represented by $(0,0,1,1,0,0,1,1)$ for $4$ electrons in $4$ spatial molecular orbitals ($8$ spin-orbitals).

Normally in NQS, the output of the network (when evaluating the value of the wavefunction, see below) is a single complex number corresponding to the natural logarithm of the wavefunction evaluated on that configuration \cite{Vicentini2022}. However, for this work it suffices to use a real-valued neural network, as discussed below.

\subsection{Initial States}\label{section:initial_states}
As with any eigenstate solver an important first step is to decide on a good initial approximation to the wavefunction. Minimally, it should contain nonzero overlap with the true ground state. In our case, the input approximation appears as training data, which is initially obtained from projective measurements of the quantum state in the computational basis, producing a set of electronic configurations. The quality of this initial state is expected to play a crucial role in determining how well the energy converges to a desired accuracy \cite{McClean2018,Bennewitz2022,Gunlycke2024}.

Many traditional ground-state solvers begin with the Hartree--Fock (HF) state, as it provides a single-determinant reference that is computationally inexpensive to obtain. As we have just discussed, the HF state is associated with a Fock space constructed from a set of molecular spin-orbitals obtained from the Hartree-Fock method, but in cases with significant static correlation, it is potentially advantageous to utilize a multi-reference approach (e.g. CASSCF). The latter would provide a significantly improved starting point by capturing the dominant configurations of the correlated wavefunction while introducing a more convenient set of molecular orbitals to construct the Fock space. 

Aside from a single reference or a sparse collection of multiple reference configurations, one could imagine the training data being derived from an MPS approximation obtained through the DMRG algorithm. The form of the MPS allows efficient sampling \cite{Ferris2012}, enabling the possibility of incorporating static-correlation effects within the trained network right away, at least in systems where DMRG can capture them well \cite{Chan2011}. On the other hand, it can be hypothesized that systems with high correlations would benefit from training data obtained as measurement data from a quantum simulation.

Ideally, the quantum computer would prepare the ground state exactly. While the construction of even an approximate ground state using VQE is not guaranteed, as mentioned in the Introduction, exact ground-state preparation \textit{is} guaranteed via adiabatic evolution. However, this is not possible on current QPU's given the short decoherence times. A more realistic solution to recovering relevant correlation effects is to, instead of slowly evolving the system over a long series of small time steps, take only a handful of larger time steps. These steps would need to be large enough to explore regions of the Hilbert space  away from the Hartree--Fock approximation and closer to the true ground state, but still small enough to limit excitations. This is the guided sampling ansatz technique of Ref. \cite{gunlycke2025}. While here we take the exact ground state as a starting point for comparison against very weak ground-state approximations, more practical implementations would benefit from utilizing these types of guiding states as effective ground-state surrogates.

Despite the potential advantages of a strong initial ground-state approximation, our numerical results presented in Sec. \ref{section:Results} indicate that the benefits of using initial states beyond Hartree–Fock are only realized for sampled subspaces sufficiently small with respect to the size of the full Hilbert space. Otherwise, single reference states appear to be an effective starting point in the systems we analyze, as demonstrated in Sec. \ref{section:Results}.

\subsection{Auto-regressive Neural Network (ARNN) Representation}\label{section:ARNN}

\begin{figure}[t]
    \centering
\includegraphics{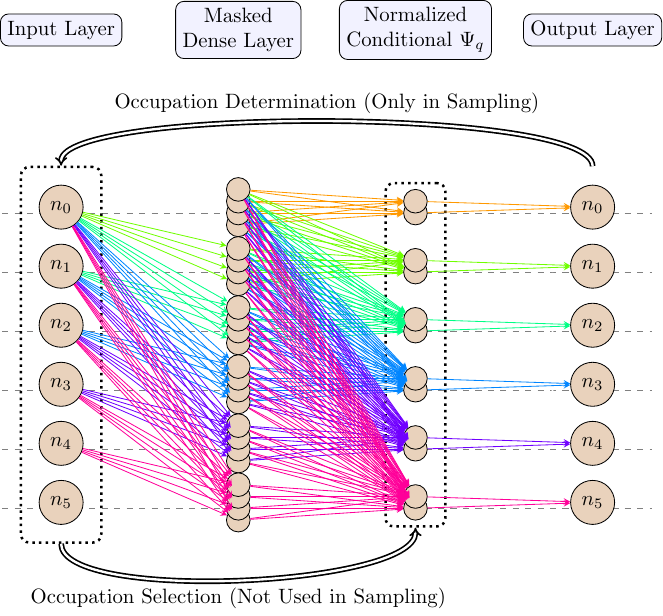}
    \caption{Auto-regressive Neural Network constructed from masked-dense layers, adopted from NetKet's ``ARNNDense" model. The depicted model acts on bitstrings of length six and uses two masked-dense layers with multiple features per bit (four for the first layer). The final masked-dense layer has two features from which binomial probabilities conditioned on preceding bits are computed. Arrows of the same color between layers correspond to all of the information fed in to one particular bit in the succeeding layer from the preceding layer, and this color is reused for the same exact bit lying at the heads of the arrows from layer to layer. In computing wavefunction values at given configurations, the information from the input configuration is fed forward to select which neuron in the second-to-last layer is kept for computing the outputs $\log\left(\Psi_q\left(n_q\right)\right)$ (not configurations as shown in the Figure), which we then sum over to compute $\log\Psi\left(n\right)$, see Eq. \eqref{eq:wvfn_output}.}
    \label{fig:our_ARNN}
\end{figure}

% \subsubsection{ARNN States}\label{section:ARNN_states}
Following Refs. \cite{Sharir2020,Barrett2022,Bennewitz2022}, we realize an efficient characterization of the ground state of a molecular system with an auto-regressive neural network (ARNN). The network we use is displayed in Fig. \ref{fig:our_ARNN}, and it is equipped with the set of real parameters (weights) $\{\alpha_k\}$, which we wish to train. While the general structure of the network is based upon the Neural Auto-regressive Density Estimator (NADE) \cite{Uria2016} and the Masked Auto-encoder for Density Estimation (MADE) \cite{germain2015}, the specific architecture is adopted from Netket's masked-dense ``ARNNDense" network structure \cite{Vicentini2022}. Masking refers to process of applying operations which hide unwanted portions of data. In the context of auto-regressive processes, it prevents ``earlier" neurons from being dependent on  ``later" ones, facilitating the use of Bayesian methods and automatically yielding a neural network that represents a normalized probability distribution.

The Figure displays an input layer, two masked-dense layers, and an output layer. Recall that $M$ denotes the number of spin-orbitals, which in this context can be referred to as bits. We see that each layer has $M$ bits, although the hidden layers contain multiple features. For example, the Figure shows four features in the first hidden layer and two in the second. The second (final) hidden layer represents a local conditional probability distribution for each bit in the sequence, normalized over the two possible states (0 and 1 bit value corresponding to a physically unoccupied or occupied spin-orbital) and conditioned on the configuration of the preceding bits. Note that in this construction we are free to choose the number of masked-dense layers, as well as the number of features per bit in every masked-dense layer except the last one, which is given the same number of features as the size of the local Hilbert space (dimension two for electrons).

For our purposes, the network we employ technically needs to only output real values, whereas in most settings like VMC a complex-valued wavefunction is essential for capturing the appropriate phase structure of the wavefunction. Here however, we use sampling only to approximate the square-amplitudes of the wavefunction, and off-load the fixation of phases to the subspace diagonalization portion (Step 5). This simplification can be thought of as a special case of VMC. As shown in Appendix \ref{appendix:method_improvements}, the energy expectation may be written as an empirical average of a modified VMC local energy estimator defined in terms of a reference wavefunction $\Psi_0$ and variations described by a function $\lambda$ of some variational parameters $\vec\theta$:
\begin{align}
    \Upsilon(\vec\theta) &\approx \frac{1}{\vert\mathcal{S}\vert}\sum_{n\in\mathcal{S}}\left[\sum_{n'}H_{nn'}e^{-i\lambda^*(n;\vec\theta)}e^{i\lambda(n';\vec\theta)}\frac{\Psi_0(n')}{\Psi_0(n)}\right]N_{n}\ ,\nonumber\\
\Lambda(\vec\theta) &\approx \frac{1}{\vert\mathcal{S}\vert}\sum_{n\in \mathcal{S}}e^{-2\text{Im}\lambda(n;\vec\theta)}N_{n}\end{align}
where $\mathcal{S}$ represents a sampled subspace of size $\vert\mathcal{S}\vert$ and $N_{n}$ denotes how many times configuration $n$ is sampled. The energy expectation is realized as $\langle H\rangle=\Upsilon/\Lambda$. Now suppose that in a quantum chemistry problem that chemical accuracy can be reached even if $\Psi$ is approximated as a function with support over some compact subspace. Then if $\mathcal{S}$ includes that subspace, we may truncate the sum inside the brackets of $\Upsilon$ to $n'\in\mathcal{S}$. This occurs when we construct $\lambda$ such that the variational ansatz has support over $\mathcal{S}$ only. This converts VMC into an exact diagonalization problem:
\begin{align}\label{eq:var_to_diag}
    \Upsilon(\vec\theta) &\approx \sum_{n,n'\in\mathcal{S}}\left(e^{-i\lambda^*(n;\vec\theta)}\sqrt{\frac{N_{n}}{\vert\mathcal{S}\vert}}\right)H_{nn'}\left(e^{i\lambda(n';\vec\theta)}\sqrt{\frac{N_{n'}}{\vert\mathcal{S}\vert}}\right)\ ,\nonumber\\
\Lambda(\vec\theta) &\approx \sum_{n\in\mathcal{S}}\Bigg\vert e^{i\lambda(n;\vec\theta)}\sqrt{\frac{N_{n}}{\vert\mathcal{S}\vert}}\Bigg\vert^2=1\end{align}
Here, $\lambda$ has been adjusted so that $\Psi_0\left(n\right) = \sqrt{\frac{N_{n}}{\vert\mathcal{S}\vert}}\ge 0$. Eq. \eqref{eq:var_to_diag} represents how sampling can fix a proper subspace for exact diagonalization, and then how amplitudes and phases are adjusted (through $\lambda$) to obtain an accurate ground-state approximation.

This amplitude-only sampling approach mirrors the ``Quantum-Selected Configuration Interaction" (QSCI) method employed in Refs. \cite{kanno2023,gunlycke2025,Moreno2025}, which measure quantum states on a quantum computer only in the computational basis. However, modifying the network to express complex wavefunctions is straightforward, see Ref. \cite{Bennewitz2022} or Appendix \ref{appendix:method_improvements}.

The full wavefunction $\Psi$ for an ARNN is given by \cite{Sharir2020}
\be\label{eq:cond} \Psi(n) = \prod_{p=0}^{M-1}\Psi_q(n_q|n_0,n_1,\ldots,n_{q-1}),\ee
where each $\Psi_q$ is a local conditional wavefunction whose absolute square is the conditional probability $P_i(n_q|n_0,\ldots,n_{q-1})$. For each $q$, the logarithms of the wavefunction for each of the two outcomes in the local probability distribution are stored in the two features (of the $q^\text{th}$ bit) in the second-to-last layer. 

From here we need to decide what calculation we are performing. If we are interested in computing the wavefunction $\Psi$ at a specific configuration $n$, then we feed forward the known orientation $n_q$ of the $q^\text{th}$ bit to select the appropriate conditional wavefunction $\Psi_q(n_q|n_0,n_1,\ldots,n_{q-1})$ and output its logarithm in the final layer. As indicated by the dashed, horizontal lines in the Figure, this selection is done from input $q$ to output $q$, and does not interfere with the auto-regressive process as this property was only necessary up to the second-to-last layer (now we are only selecting a probability amplitude from an already-normalized distribution). Thus the extracted wavefunction value at configuration $n$ is
\be \label{eq:wvfn_output}\Psi\left(n\right) = \text{exp}\left[\sum_{p=0}^{M-1}\log\{\Psi_q\left(n_q\right)\}\right]\ , \ \ \Psi_q(n_q)\equiv \Psi_q(n_q|n_0,n_1,\ldots,n_{q-1}).\ee

Instead of computing the wavefunction at a pre-determined orientation, the situation may require us to sample the orientations directly using the same network, exploiting its generative capability. In this case, each output neuron yields a bit value rather than a conditional wavefunction value, and these values are determined by sampling, in succession, each local conditional probability distribution in the second-to-last layer. Every time a bit value is determined, we use that as input for another forward pass through the network to determine the value of the next bit. A fast way to generate samples in this way is reviewed in Sec. \ref{Section:Sampling}.

The localized conditional distributions are realized using a masking procedure. The first (top) output bit does not depend on any of the inputs; it only depends on the parameters $\{\alpha_k\}$. The second depends only on the first input bit $n_0$ and the $\{\alpha_k\}$. The following output bits depend only on the previous input bits and the $\{\alpha_k\}$, and we note that none of them depend on the final input bit. The mask itself is realized as a binary triangular matrix multiplied element-wise by the dense weight matrix connecting adjacent layers. 

% A ``mask" is also applied between the two final layers to account for the orientation selection, meaning only one of the two arrows reaches the last layer, per bit. If computing wavefunction values, this mask depends on the input configuration that is queried. If sampling, it depends on the sampling result from each of the conditional distributions $P\left(n_q|n_0,\ldots,n_{q-1}\right)=\vert\Psi_q(n_q)\vert^2$.

\subsection{Training}

To train the neural network to approximate the ground-state wavefunction, we begin with a dataset of configurations sampled from an external approximation, as described in Sec. \ref{section:initial_states}. Each configuration in this dataset corresponds to a bitstring representing the occupation pattern of spin-orbitals. The network is trained by minimizing a cost function known as the Kullback–Leibler (KL) divergence \cite{Kullback1951}. It quantifies the difference between a probability distribution $P(n)$ and a model distribution, $P_{\vec\alpha}(n)$, parameterized by a set of real numbers $\{\alpha_k\}$. The KL divergence is defined as
\begin{equation}\label{eq:KL_div}
D_{\text{KL}}(P \,\|\, P_{\vec\alpha}) \equiv \sum_{n} P(n) \log \frac{P(n)}{P_{\vec\alpha}(n)}.
\end{equation}

Keeping in mind that $P(n)$ does not depend on the network parameters, we may estimate the KL divergence and its gradient using the empirical averages
\begin{gather}
D_\text{KL}(P_{\text{data}} \,\|\, P_{\vec\alpha}) \approx \frac{1}{\vert\mathcal{S}\vert}\sum_{n\in\mathcal{S}} \log \frac{P_\text{data}(n)}{P_{\vec\alpha}(n)} \nonumber\\
 \partial_{\alpha_k} D_\text{KL}(P_{\text{data}} \,\|\, P_{\vec\alpha}) \approx -\frac{1}{\vert\mathcal{S}\vert}\sum_{n\in\mathcal{S}}\partial_{\alpha_k} \log P_{\vec\alpha}(n),\label{eq:KL_grad}
\end{gather}
where $\mathcal{S}$ is a (mini-)batch of size $\vert\mathcal{S}\vert$ collected from the dataset. In minimizing $D_\text{KL}$, one maximizes the likelihood that $\mathcal{S}$ was obtained directly from the model distribution $P_{\vec\alpha}$. The definition itself indicates that proposed models lacking vital peaks occurring in $P_\text{data}$, and by extension $P$, are punished in the training process. 

Because we use an auto-regressive neural network (ARNN), the model outputs a normalized probability distribution over bitstrings by construction. Eqs. \eqref{eq:cond} and \eqref{eq:wvfn_output} imply the factorization of a configuration probability into conditional probabilities, allowing us to compute exact log-probabilities of each configuration in the training set directly. Without this, one cannot compute $\log P_{\vec\alpha}$ and its derivatives directly as seems to be required by Eqs. \eqref{eq:KL_grad}. RBM's --- used in Ref. \cite{Herzog2023} --- do have associated with them an analytical expression for the derivatives, but this expression involves an average over the model distribution as well as the data distribution. Computing the former requires the generation of Markov chains, which generally suffer from auto-correlation effects if the chains are too short \cite{MuellerKrumbhaar1973,Wu2021}. 

Since the summation in Eq. \eqref{eq:KL_grad} is over configurations sampled from a dataset, we need only evaluate the derivative of the log-probabilities assigned by the model to the configurations provided in this dataset (similar to the first type of calculation described in Sec. \ref{section:ARNN}). That is, there is no need to perform auto-regressive sampling during the training. This contrasts with VMC where we would have to generate samples from the network directly to compute gradients within each training step. Any auto-regressive sampling is performed outside of the training loop (to be detailed next).

This training procedure allows the ARNN to learn the approximate structure of the distribution associated with ground-state Born probabilities directly from the approximate ground state $\ket{\Psi_i}$ by maximizing the likelihood that the samples from $\ket{\Psi_i}$ were generated from a distribution described exactly by the ARNN. By not directly using the wavefunction $\ket{\Psi_i}$ in conjunction with the non-empirical Eq. \eqref{eq:KL_div}, we avoid exactly fitting the truncated approximation and instead aim for an improved representation of the exact ground state, capturing configurations which likely contribute significantly to the exact GS but did not appear in $\ket{\Psi_i}$. This is made possible through the generalization properties of NN's. No knowledge of the underlying Hamiltonian is required during this training step of the algorithm, again differentiating itself with the VMC approach. The minimization of the cost function is achieved using the ADAM optimizer \cite{kingma2017} which is itself a form of Stochastic Gradient Descent (SGD) \cite{bottou2018}. 

% We make one final observation from Eq. \eqref{eq:KL_grad}. The accuracy of these approximations depends on the variances of both the data and its model over the various configurations $n$ appearing in the data distribution. If it is in our capacity to deform the incoming data itself to reduce these variances, then we have no need for extraordinarily large batch sizes. We re-visit this below.

\subsection{Sampling}\label{Section:Sampling}
Once we have obtained a neural-network approximation to the computational basis probabilities suggested by the training data, we now have an efficient classical object that we have considerable control over. However, the probabilities obtained from the network may yet still be a very poor reproduction of what would be observed from sampling the actual quantum state. This has been shown in Ref. \cite{Bennewitz2022}, where attempts to reproduce the quantum state using Neural Quantum State tomography \cite{Torlai2018} still results in very low fidelity, and therefore one must fix the amplitudes and phases using VMC. In our case, we only need a rough sketch of the Born probabilities to help identify which subspace of the full Hilbert space we need to achieve a far better approximation to the ground state. This requires us to analyze the techniques we use to sample the neural network to ensure that sufficient guidance can be provided.

\subsubsection{Handling un-physical States} \label{Section:sampled_symmetries}
The ARNN we use for performing SCI could be such a bad representation of the sampled initial state that it can and does have non-negligible support outside of the physical subspace imposed by system symmetries. In principle, additional training data and more expressive NQS's could learn that this support should not occur, but our aim is to reduce computational cost. Another option is to enforce symmetries by hand, and this has been done in some works \cite{Barrett2022,malyshev2023}. 

It turns out the latter is not necessary in our case. As our goal is to find the most relevant configurations, it does not matter where the distribution has its tails, as these regions are to be excluded from the calculation anyway. In fact, it is perhaps advantageous to keep the full Hilbert space, as it gives the network more flexibility in the training process (as long as the relevant states keep considerable support, we are allowed to ``rotate" the tails outside of the physical subspace). Ref. \cite{malyshev2023} came to a somewhat similar conclusion, warning that constraints from imposing symmetries could affect network expressivity, e.g. all of the pink connections in Fig. \ref{fig:our_ARNN} would lose their variational properties in ensuring that the final bit has the correct orientation.

If the support outside of the physical subspace is reasonably small, then simply discarding un-physical states should be as good as if not better than enforcing symmetries, even if the collective contribution from these states make up a majority of the samples. Indeed, it was found that enforcing electron-number conservation by hand did not result in significant improvements in the tests leading up to the results displayed in the next Section, and therefore we opted to simply discard un-physical states. This does not mean there isn't room for improvement. For instance, the ``configuration recovery" strategy outlined in Ref. \cite{Moreno2025} may prove quite effective in future work, as it provides a procedure to convert already-sampled faulty states into meaningful ones. In any case, for our tested systems, we discarded any configurations with the incorrect electron number or belonging to the incorrect representation of the relevant point group. We also require that the number of spin-up electrons is equal to the number of spin-down electrons. Both of these are straightforward to evaluate, as they involve simply counting orbital occupations and noting to which Abelian subgroup representation each occupied orbital wavefunction belongs (the many-electron GS wavefunction should be entirely symmetric under all subgroup operations).

\subsubsection{Temperature Scaling}
Without further modification, our algorithm would have two important weaknesses. First, when the size of the training set is small, the neural network in Step 3 tends to be over-fitted to the training data provided, making it difficult to identify additional important configurations that were not seen during training. Regularization techniques such as $L_2$ regularization and dropout \cite{neupert2022} are expected to improve generalization, and in our implementation, dropout is applied effectively during training. However, dropout alone only goes so far in enabling the discovery of unseen configurations that are relevant to the ground state.

The second limitation concerns the way in which we sample from the trained auto-regressive neural network. While the ARNN enables efficient sampling of the corresponding probability distribution over configurations, this sampling is performed with replacement. This is not ideal in our setting, where we are interested primarily in discovering a diverse set of unique configurations that make significant contributions to the ground state, rather than repeatedly sampling the same dominant configurations. In an ideal scenario, we would sample without replacement, but this is not straightforward under our current setup. However, we can approach this behavior by modifying the sampling distribution to reduce the dominance of highly probable states and boost the likelihood of sampling initially improbable, yet still relevant, configurations.

One technique for this is known as temperature scaling, which involves reshaping the distribution by raising the probabilities to the power of an inverse temperature parameter $\beta$. After proper renormalization, this suppresses over-dominant configurations while increasing the probability of sampling (formerly) lower-weight, yet important, configurations. In systems small enough to compute the exact probabilities $P(n)$, this can correspond to
\be\label{eq:sparse_rescale}
P(n) \to \frac{P(n)^\beta}{\sum_{n'} P(n')^\beta}, \quad \beta > 0.
\ee

However, exact enumeration of the configurations becomes intractable in larger systems. Fortunately, ARNNs allow temperature scaling to be implemented locally within their generative process: we apply the scaling independently to each conditional probability in the chain rule expansion,
\be\label{eq:ARNN_rescale}
P_q(n_q \mid n_0, \ldots, n_{q-1}) \to \frac{P_q(n_q \mid n_0, \ldots, n_{q-1})^\beta}{\sum_{n_q'} P_q(n_q' \mid n_0, \ldots, n_{q-1})^\beta}\ ,\ \ P_q=\vert \Psi_q\vert^2
\ee
It is straightforward to show that this change, which is applied locally over all sites, still preserves normalization over the $n$ (note that, despite having similar effects, the global distribution induced by Eq. \eqref{eq:ARNN_rescale} is not identical to that given by Eq. \eqref{eq:sparse_rescale}). Eq. \eqref{eq:ARNN_rescale} enables efficient control over the effective entropy of the sampled-state distribution without sacrificing sampling speed or requiring expensive Markov Chain Monte Carlo (MCMC) methods. In contrast, generic neural-network wavefunctions that lack auto-regressive structure require global rescaling and would rely on MCMC to get samples from the modified distribution. Moreover, quantum hardware lacks the capability to deform the probability distribution associated with prepared quantum states altogether, leaving the discovery of configurations through QSCI an inefficient process. Temperature scaling via ARNNs sidesteps these difficulties entirely.

\begin{figure}
    \centering
    % \subfigure[$\ \beta=1$]{\includegraphics[width=0.65\linewidth]{figs/C2H2_probs_alpha_1.pdf}}
    % \subfigure[$\ \beta=0.75$]{\includegraphics[width=0.65\linewidth]{figs/C2H2_probs_alpha_075.pdf}}
    % \subfigure[$\ \beta=0.4$]{\includegraphics[width=0.65\linewidth]{figs/C2H2_probs_alpha_04.pdf}}
    % \subfigure[$\ \beta=0.2$]{\includegraphics[width=0.65\linewidth]{figs/C2H2_probs_alpha_02.pdf}}
    \includegraphics[width=0.85\linewidth]{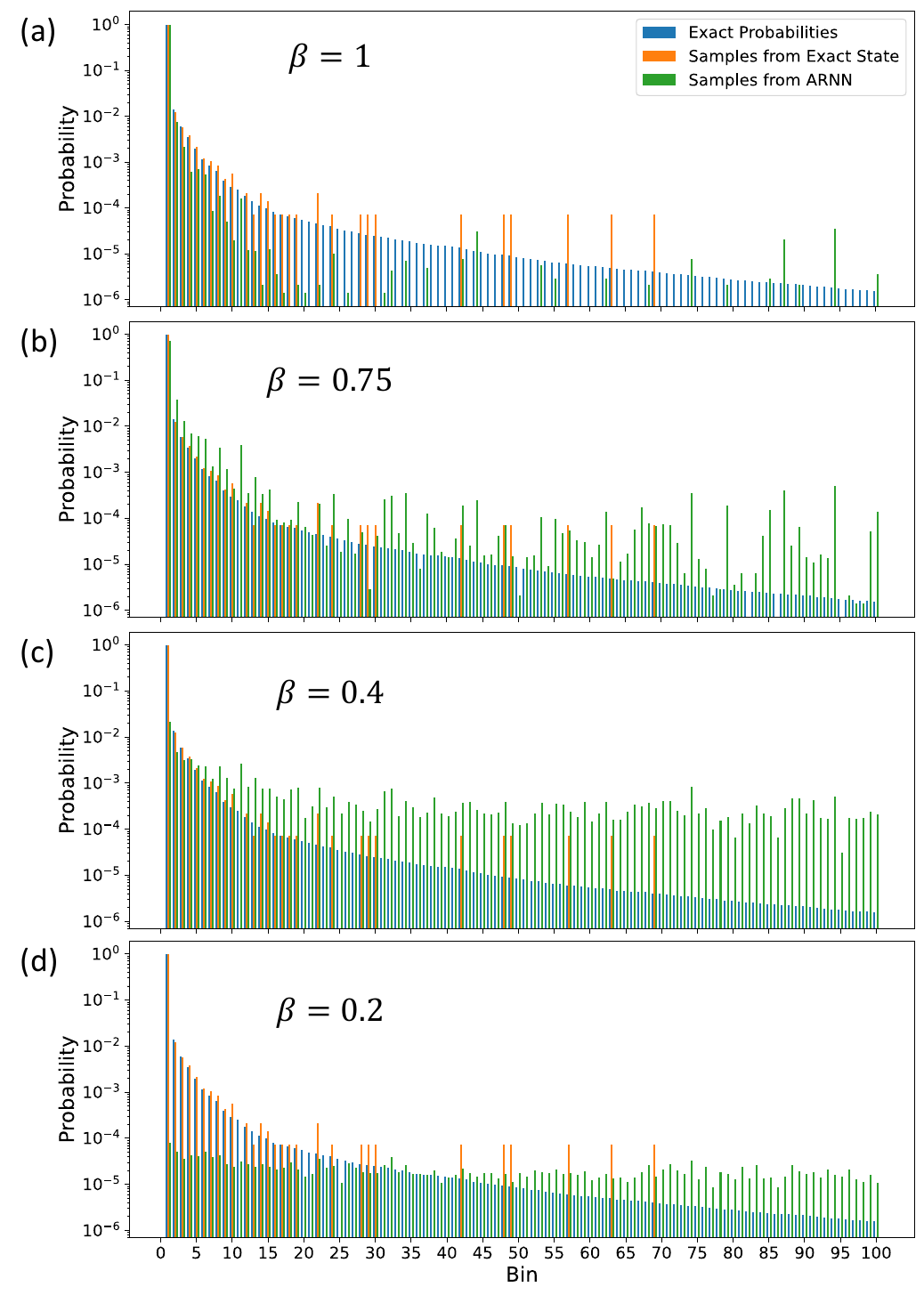}
    \caption{Modified ARNN probability distribution for various values of inverse temperature $\beta$, for the $\text{C}_2\text{H}_2$ molecule considered in Section \ref{section:Results}. The horizontal axis organizes, into bins of $20$ configurations, the first $2000$ configurations that appear when sorting them based on the absolute squares (Born probabilities) of the exact wavefunction amplitudes. The blue bars correspond to the weights of each bin, for the exact GS. The orange bars are derived from direct sampling of the exact Born probabilities associated with the blue bars, with a sample size of $N_\text{T}=1.4\times 10^4$. That is, the orange weights are empirical, while the blue weights are exact. The green bars also represent empirical weights, but this time take a sample size $N_\text{N}=1.4\times 10^6$ from the ARNN. However, the ARNN itself is trained on only $N_\text{T}$ samples of the GS.}
    \label{fig:distribution_flattening_example}
\end{figure}

A visual of the effect of the parameter $\beta$ is shown in Fig. \ref{fig:distribution_flattening_example}, where we identify the $2000$ highest-contributing configurations for the $\text{C}_2\text{H}_2$ molecule, according to their contribution to the exact GS, and observe their corresponding weight under both direct ground state and ARNN sampling. The configurations are organized in to bins containing twenty configurations each. Although more samples $N_\text{N}$ are drawn from the ARNN than the number of samples $N_\text{T}$ drawn from direct GS sampling, to build the sampled subspace, the ARNN is trained only on $N_\text{T}$ GS samples. We see that at $\beta=1$ [panel (a)] the ARNN does an excellent job at matching true probabilities for the most dominant contributions to the ground state. For less dominant but relevant contributions, the network is still able to assign non-zero probabilities to important states, yet they are so small that a large number of samples is required to notice them. Therefore, we rely on lower values of $\beta$ [panels (b) and (c)] to bolster their probabilities by orders of magnitude to make them more visible. We see that one can find many important states that direct sampling (orange bars) did not reveal. This comes at the cost of some mismatched probabilities for the dominant contributions, but for a wide range of values for $\beta$ this is not an issue, as the diagonalization step does not care about the multiplicity of each configuration in the sampled subspace. Panel (c) shows that $\beta=0.4$ is close to ideal, as it reveals many configurations while still preserving the ordering according to how much each state contributes to the GS. In panel (d) we see what happens when $\beta$ is too small -- the distribution is so close to uniform that our ability to find physically relevant configurations is effectively removed. Indeed, panel (d) starts to look more like random sampling.

% Thus far we have only considered temperature scaling implemented during sampling, not training. This allows us to preserve a faithful representation of the learned ground state approximation while expanding the sampled subspace more broadly and adaptively. 

For each iteration of our SCI algorithm, we fix the total number of samples $N_\text{N}$, and tune the inverse temperature $\beta$ until the number of unique configurations $N_\text{U}$ meets a specified value --- or their quality starts to degrade compared to larger $\beta$. That is, flexible exploration is encouraged for small values of $\beta$, but if $\beta$ becomes too small --- approaching zero --- the distribution tends toward uniform, resulting in random exploration. As $\beta \to 1$, the original learned distribution is recovered. Thus, temperature scaling provides direct control over sampling diversity without inflating sample sizes.

Recall that ``iteration zero" is seeded by an externally provided ground-state approximation. However, afterward, we seed the protocol with a sparse approximation from the previous iteration. This permits us to easily adopt the ideas of temperature scaling to the training step as well, using Eq. \eqref{eq:sparse_rescale}, and it proves to be even more effective than when it is applied to sampling. Specifically, a separate inverse temperature $\beta_0$ can be used to reshape the distribution associated with the sparse approximation $\ket{\Psi_i}$, and results are noteably improved if we set $\beta_0\approx 0.4$ for all iterations beyond iteration zero. We assume below that this has been applied in all of our results. Thus, in Sec. \ref{section:Results}, we reserve the term ``temperature scaling" for the sampling parameter $\beta$, and not the constant value $\beta_0=0.4$ applied to $\ket{\Psi_i}$ during training.

% and the network used to train it. We leave $\beta$ as a free parameter to control diversity during subspace expansion in the current iteration. In this way, we retain flexibility between approximating the true ground state and discovering new Fock states likely to improve the subspace basis.

\subsubsection{Fast Auto-regressive Sampling}
The auto-regressive architecture utilized for this work has the benefit of being sampled in a very fast manner according to the procedure described in Ref. \cite{Barrett2022}. Instead of taking a desired sample size $N_\text{N}$ and generating a separate bitstring for each sample, we begin by associating all of the samples with one empty bitstring and then grow the bitstring bit-by-bit, keeping track of what unique sub-strings are generated and also what proportion of the samples become associated with each of these sub-strings. This is achieved by sampling from a binomial distribution specified by the local conditional probabilities $P_q$. Since we only need to condition these binomial distributions on what bitstrings were already sampled, this sampling technique only scales with the number of unique bitstrings, not $N_\text{N}$, since for each bit we only need to evaluate the network across a batch containing just one copy of each unique bitstring. 

Combined with temperature scaling, we thus have a way of generating a large number of relevant bitstrings in the sampling step. In the cases in Sec. \ref{section:Results} where temperature scaling is utilized (during sampling), it is implemented for only one to three iterations, after which it ceases to be effective. After this point, fast auto-regressive sampling becomes the dominant, necessary method to generate large sample sizes $N_\text{N}$ for the continued improvement of the energy estimate.

\subsection{Subspace Diagonalization and Iteration} 
Assuming that we have a relatively small number of unique configurations $N_\text{U}$, we want to treat the Hamiltonian in the subspace formed by these states. For smaller systems, it is possible to obtain the Hamiltonian matrix, in which case we can simply pick out the rows and columns corresponding to the states of interest. For larger systems, it is not possible to form the matrix in the full Hilbert space, and therefore we focus instead on identifying the matrix elements directly in the subspace, which for many systems is possible given that the molecular Hamiltonian has polynomially many terms.

% If we are interested in keeping states we have already sampled through multiple iterations of our full procedure, then it is important that we remember the matrix elements between sampled states and ALL Fock states they connect with, even if those connected states were not yet sampled, as they may be sampled in future iterations. At the end of each iteration, we are free to temporarily shave off not-yet-sampled states.

% \subsection{Iteration}

We have indicated that it is possible to take the result of subspace diagonalization and use this quantum state $\ket{\Psi_i}$ to initialize the following iteration. This is due to the fact that this quantum state is represented by a sparse vector in the full Hilbert space, with only $N_\text{U}$ non-zero elements. This means that sampling the new approximation in the computational basis is easy as long as $N_\text{U}$ is kept under control, allowing us to freely adjust the number of training samples $N_\text{T}$ within each iteration if needed. 

As discussed, an efficient representation of the ground-state approximation allows us to enhance the procedure by applying temperature scaling to the training data ($\beta_0$ temperature scaling). This is realized by applying Eq. \eqref{eq:sparse_rescale} to the obtained sparse approximation. This effectively smooths out the distribution we are training against and lays out more clearly the relationships between the relevant amplitudes, which are hard to observe without re-weighting. We have found that this does in fact expose additional relevant configurations for consideration in the subspace diagonalization. 

In Section \ref{Section:Sampling} we noted that Eq. \eqref{eq:KL_grad} contains empirical averages computed over the support of the dataset only, which itself is contained in the support of $\ket{\Psi_i}$, not the network model. The situation is different in both VMC \cite{Trail2008} and the RBM training of Ref. \cite{Herzog2023}, where the training requires taking averages over the model distribution. As $\ket{\Psi_i}$ is sparse and also made smooth via $\beta_0$, both the required model complexity and necessary number of training samples $N_\text{T}$ are reduced. As our approximation improves, then in later iterations we need to both switch to more complicated models as well as increase $N_\text{T}$ accordingly, and this is done in our demonstrations.

\section{Results} \label{section:Results}
In this Section we apply our iterative algorithm to a few small molecules to demonstrate its effectiveness. For the larger molecules acetylene ($\text{C}_2\text{H}_2$) and ethelyne ($\text{C}_2\text{H}_4$), we restrict ourselves to a minimal basis, STO-3G, whereas for the smaller molecules $\text{H}_2\text{O}$ and $\text{C}_2$, we explore the larger basis set 6-31g \cite{Hehre1972}. As discussed in Sec. \ref{Section:Sampling}, we post-select configurations corresponding to the correct particle number and spatial symmetries so that chemical accuracy is shown to be achieved with relatively light resources. As is done in Ref. \cite{Herzog2023}, we simplify the approach by considering only the Abelian subgroups of the point groups relevant for each molecule, that is, we take $C_{2v}$ for the water molecule and $D_{2h}$ for the others. Checking electron number and point-group representation is straightforward when working in the Fock basis, as mentioned in Sec. \ref{Section:Sampling}.

\begin{figure}[t]
    \includegraphics[width=0.95\linewidth]{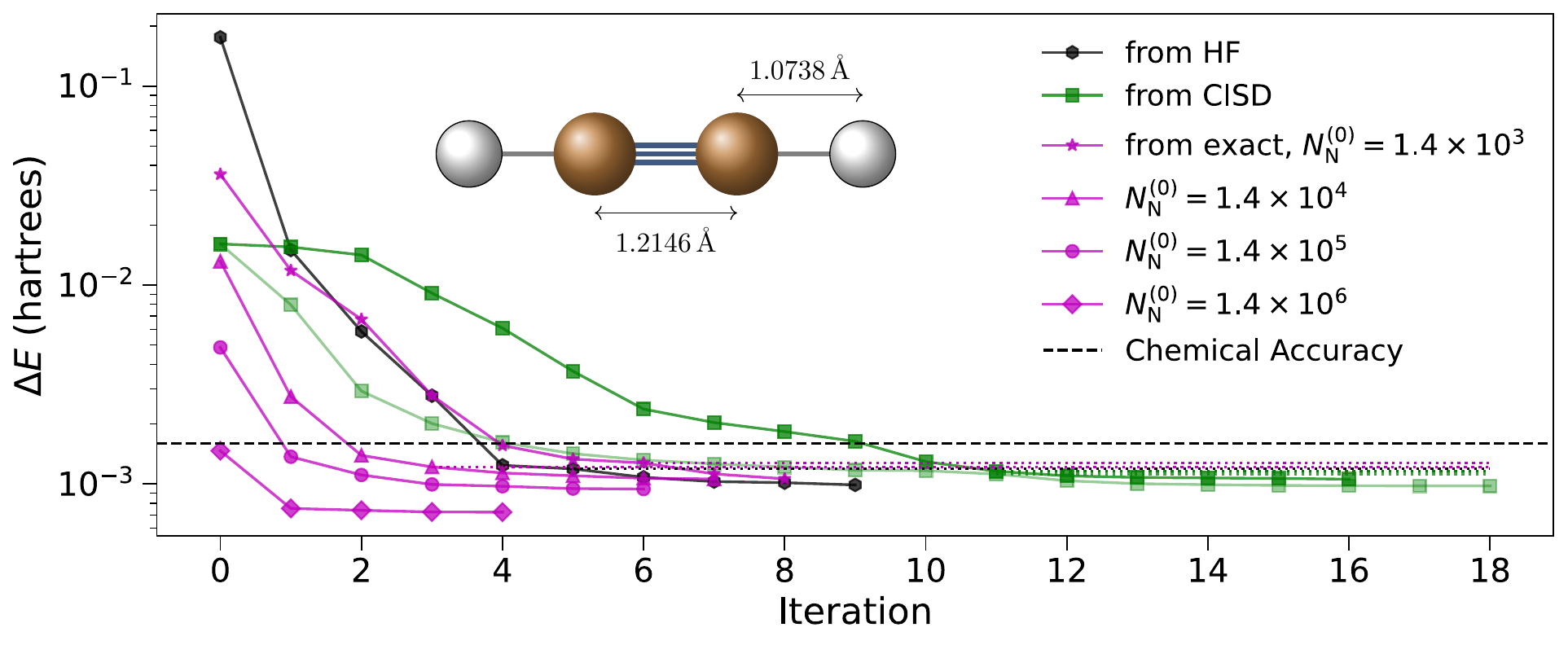}
    \caption{Ground state energy error $\Delta E$ \textit{vs} iteration number obtained for the $\text{C}_2\text{H}_2$ molecule using the STO-3G basis set, with displayed molecular geometry the same as that from Ref. \cite{Joseph2010}. The solid/faded, green curve corresponds to the raw/temperature-scaled CISD-initiated curve. When sampling from the exact ground state, we consider four cases where the number of shots differ by an order of magnitude. We take network sample size $N_\text{N}=1.4\times 10^6$ and subspace dimension $N_\text{U}=1600$ in all curves. The horizontal dashed lines in all but the bottom two curves indicate a switch in the network architecture, where the number of layers and features are doubled, and $N_\text{T}$ increases from $10^4$ to $10^5$. Only the larger model is used in the bottom two curves. Additional hyper-parameters: two masked-dense layers, four features per bit, and dropout rate 0.05 in the smaller model, each of which are doubled in the larger model. ADAM learning rate 0.001. The temperature-scaled CISD curve takes $\beta=0.4$ for the first iteration. HF takes $\beta=0.1,0.6$ for the first two iterations.}
    % , with $N_N=10^6$ and $N_U=2,000$. The two magenta ``Exact State" plots take $N_T=10^4$ and $N_T=10^6$, respectively. The temperature scaling $\beta_0=0.4$ is applied to the initial state after the first iteration, while the ARNN temperature scaling $\beta$ is varied in early iterations to encourage rapid energy minimization. Horizontal, dotted lines originate at final points where no sampled states are forcibly retained (see text for details). Additional hyper-parameters: four masked-dense layers, eight features per bit. Dropout rate 0.1 and ADAM learning rate 0.001.}
    \label{fig:convergence_vs_cheat_C2H2_10_6}
\end{figure}

Unlike Ref. \cite{Herzog2023}, our goal is to demonstrate that under strict maximum-subspace-dimension requirements, it is advantageous to train our network against the probability distribution of a given quantum state. We also illustrate the advantage of applying temperature scaling when given a weak approximation or limited GS samples. To see the advantage for the first point, we define a procedure which we illustrate with the $\text{C}_2\text{H}_2$ molecule in the STO-3G basis. See Fig. \ref{fig:convergence_vs_cheat_C2H2_10_6}. This system has $24$ spin-orbitals and $78,992$ configurations satisfying both electron-number and spatial symmetries. Taking for now the given quantum state to be the exact ground state, we first identify the minimum number of configurations needed to form a Hamiltonian subspace yielding a ground-state approximation that achieves chemical accuracy (CA). Namely, we sort the configurations according to their contribution to the exact ground state and take only the first $N_\text{CA}$, enough to get us within chemical accuracy. For our example system (with the chosen geometry) we have $N_\text{CA} \approx 800$.

The expected number of samples from the ground-state probability distribution we need to take to see all $N_\text{CA}$ states is the inverse of the probability of sampling the $N_\text{CA}^\text{th}$ configuration. For $\text{C}_2\text{H}_2$ we need about $1.4\times 10^6$ samples to recover the $N_\text{CA}=800$ most important configurations, and thus we take this to be the number of samples $N_\text{N}$ we extract from the neural network. The number of samples $N_\text{N}^{(0)}$ of the exact ground state we use for the ``iteration zero" described in Sec. \ref{section:Description_of_Algorithm} is decided with respect to this value, and would ideally be small compared to $N_\text{N}$ given the ease and speed at which we can take large sample sizes from an auto-regressive model (see Sec. \ref{Section:Sampling} and Ref. \cite{Barrett2022}). As shown in the Figure, we might take $N_\text{N}^{(0)} = 1.4\times 10^{3},\ 1.4\times 10^{4},\ 1.4\times 10^5,$ or even $1.4\times 10^6=N_\text{N}$.

Finally, we must decide on how many uniquely sampled configurations $N_\text{U}$ to keep for exact diagonalization (ED). Taking $N_\text{U}=N_\text{CA}$ is too restrictive and requires highly expressive and extremely well-trained NN's, even if the goal is just to reach chemical accuracy. Meanwhile, large $N_\text{U}$ is too expensive for ED. We therefore take $N_\text{U} = 2\times N_\text{CA}$, which is $1,600$ for $\text{C}_2\text{H}_2$. 

In the Figure, we keep applying the iterative procedure until the energy stops decreasing. When starting with the Hartree--Fock (HF) or Configuration Interaction Singles and Doubles (CISD) approximation (taking them to be $\ket{\Psi_\text{init}}=\ket{\Psi_0}$), or starting with a small sample size $N_\text{N}^{(0)}$ from the exact ground state (to generate $\ket{\Psi_0}$ from $\ket{\Psi_\text{init}}=\ket{\text{GS}}$), it is advantageous to train a smaller network and then switch to a larger network when the energy stops decreasing the first time around. The lowest energy achieved by the smaller network is indicated by the horizontal dotted lines in the Figure, which originate at the final iteration in which the smaller NN is trained. The larger network structure has twice as many layers and features as the smaller network and takes in $10$ times as many training samples $N_\text{T}$.

The fact that the energy appears to stop decreasing only after a handful of iterations is simply an artifact of highly restrictive calculational settings and, more importantly, that choosing which states to pass through to exact diagonalization is still arbitrary, in the sense that we still have flexibility in manually incorporating known relevant configurations and removing known irrelevant configurations, regardless of sampling results. Here, we make a specific, minimal choice in which the only configurations forced to contribute to the ED calculation, other than the sampled ones, are those which contribute to the CISD approximation. That is, we ensure that the CISD configurations, many of which are relevant to the ground state, appear alongside the sampled ones in case any are missed. Further energy minimization in later iterations may be carried out, for example, by cleverly forcing more and more sampled configurations to remain in the ED calculation, regardless of whether they continue to be sampled every time. As this level of fine-tuning is not our immediate objective and similarly affects all curves appearing in the Figure, we limit the calculation to our minimal choice, for each application of ED. 

In turning to the qualitative features of Fig. \ref{fig:convergence_vs_cheat_C2H2_10_6}, we first note that there are two curves corresponding to the case in which we start with the CISD state --- one case without using temperature scaling, which we call the ``raw" case, and one case where temperature scaling is applied. Additionally, we include a curve initiated by just the Hartree--Fock (HF) state, which is only possible due to temperature scaling as the trained network assigns non-zero yet incredibly small probabilities to all other configurations. We see that under the choice of calculational procedure described thus far, it can be difficult to achieve a desired accuracy when beginning from the HF and CISD approximations, especially in the raw case for the latter, whereas taking increased sample sizes from the exact ground state gives us a better edge.

There are additional striking features of the Figure. First, the raw CISD curve appears to take far longer to converge than the ones utilizing ground-state sampling. Second, even when achieving energies which lie below those of earlier iterations of the GS-sampled curves, the CISD curve does not receive a sudden speed-up in convergence, indicating that its ground-state approximations still lack vital features already seen in the GS-sampled curves. In terms of the value of the final energy reached (that is, putting aside any consideration of the time it takes to get there), the CISD curve remains competitive with the curves corresponding to $N_\text{N}^{(0)} = 1.4\times 10^3$ and $N_\text{N}^{(0)} = 1.4\times 10^4$, but at the same time exhibits a gap when compared against the curves corresponding to higher numbers of samples from the ground state, $N_\text{N}^{(0)} = 1.4\times 10^5$ and $N_\text{N}^{(0)} = 1.4\times 10^6$. 

\begin{figure}[t]
    \centering
    \includegraphics[width=0.95\linewidth]{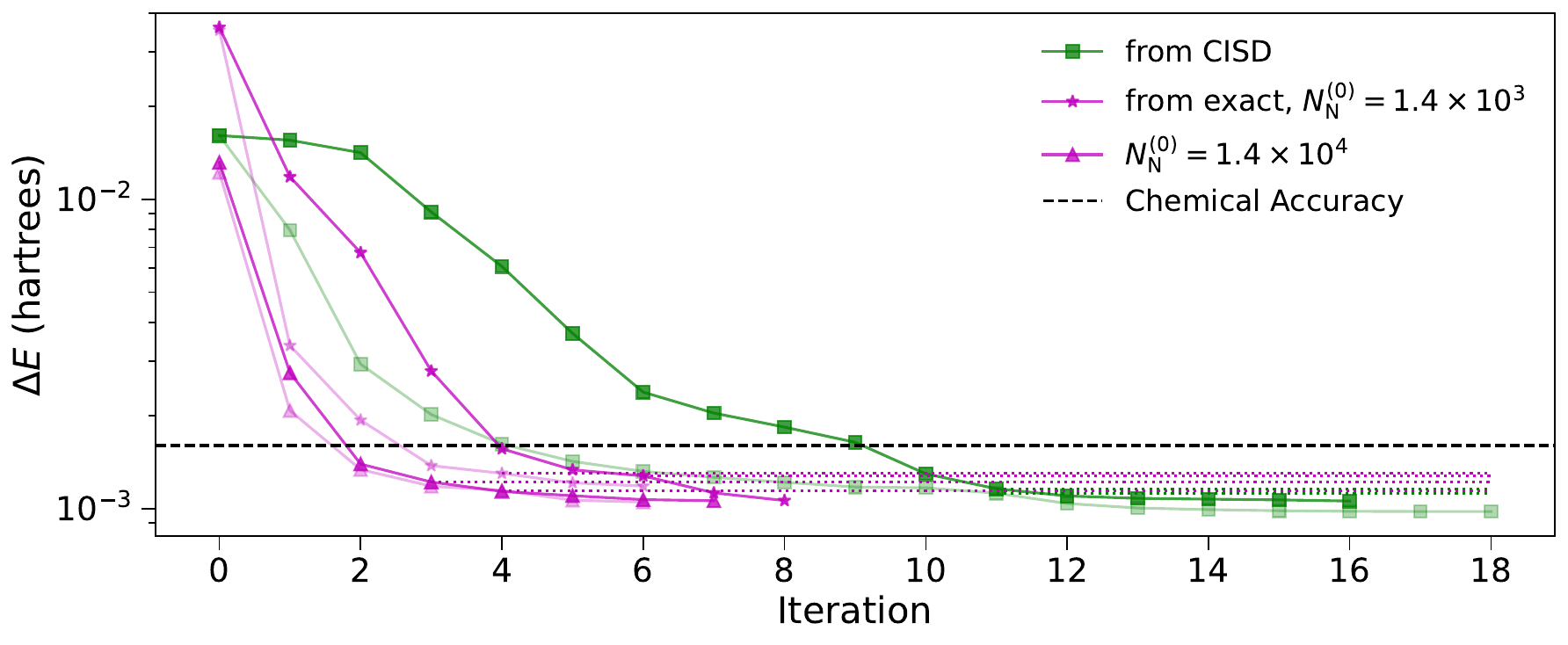}
    \caption{Fig. \ref{fig:convergence_vs_cheat_C2H2_10_6} again, keeping only the two CISD curves as well as the $N_\text{N}^{(0)}=1.4\times 10^3$ and $N_\text{N}^{(0)}=1.4\times 10^4$ curves, adding temperature-scaled versions (faded) for the two GS-sampled curves. In each case $\beta\ne 1$ is useful for up to a maximum of three iterations. The temperature-scaled CISD curve has $\beta=0.4$ for the first iteration, while the GS-sampled curves only ever go down to $\beta=0.8$.}
    \label{fig:C2H2_ts}
\end{figure}

Finally, we find that employing temperature scaling leads to a very significant speedup in the CISD curve, with even a small improvement in the final energy accuracy. In fact, the only change we made here was to set $\beta=0.4$ for the first iteration. Additionally, the HF curve (also temperature-scaled) demonstrates a very rapid approach to the accuracy regime populated by the other curves, in this case performing better than both the CISD curves. We note, however, that these temperature-scaled curves still do not compete with those corresponding to large GS sample sizes $N_\text{N}^{(0)}$. This raises the question what the effect of temperature scaling is when applied to the GS-sampled curves. We address this in Fig. \ref{fig:C2H2_ts}, where temperature scaling is applied to the curves with lower sample sizes, as temperature scaling loses its utility for larger sample sizes. Like the CISD curve, the one with the lowest $N_\text{N}^{(0)}$ received a major boost in convergence speed, whereas the curve for $N_\text{N}^{(0)}=1.4\times 10^4$ remained generally unchanged, except that a more accurate energy was achieved in the first iteration. With these considerations in mind, additional results employing temperature scaling are shown only for HF and CISD curves, so that the utility of temperature scaling (in the sampling stage) is demonstrated without complicating the displayed results.

\begin{figure}[t]
    \centering
    \subfigure[\ $\text{C}_2\text{H}_2$ iteration $i=0$]{\includegraphics[width=0.45\linewidth]{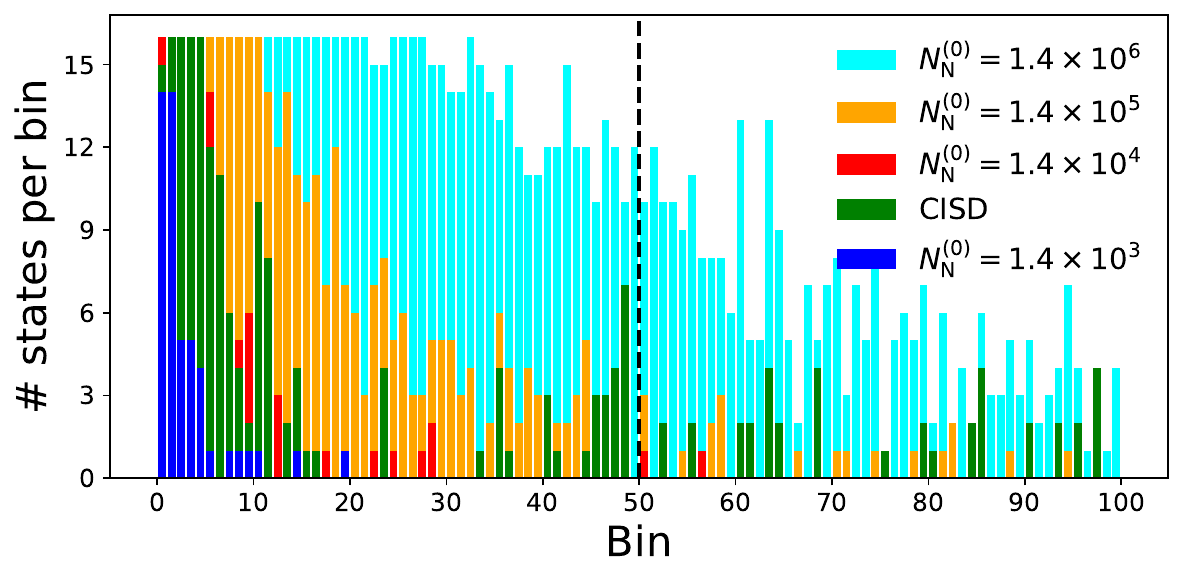}}
    \hspace{2em}
    \subfigure[\ $\text{C}_2\text{H}_2$ final iteration]{\includegraphics[width=0.45\linewidth]{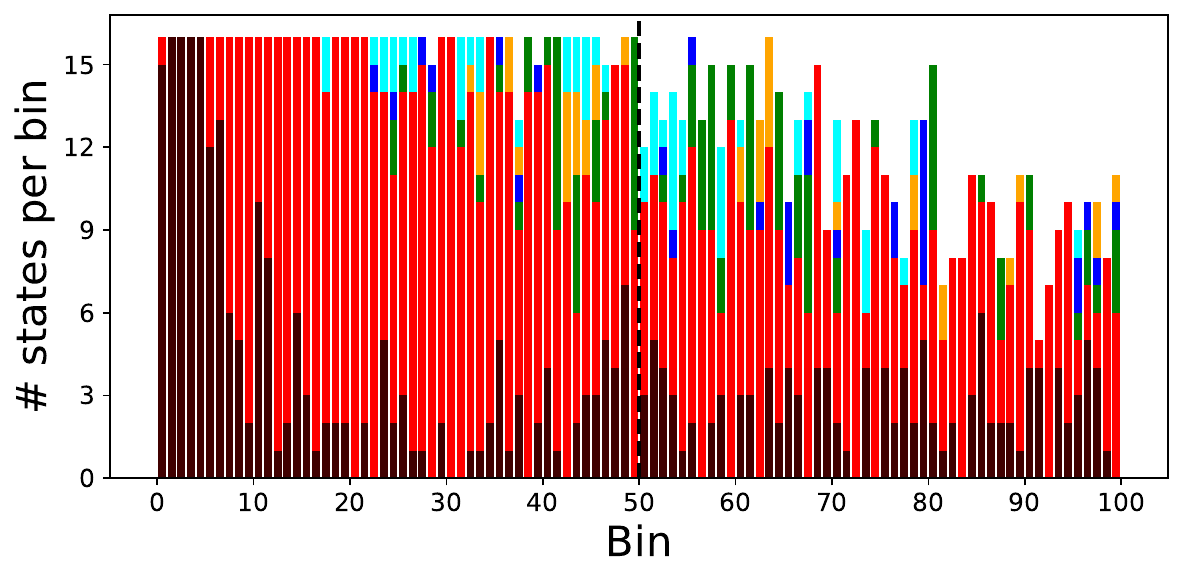}}
    \subfigure[\ $\text{C}_2\text{H}_4$ iteration $i=0$]{\includegraphics[width=0.45\linewidth]{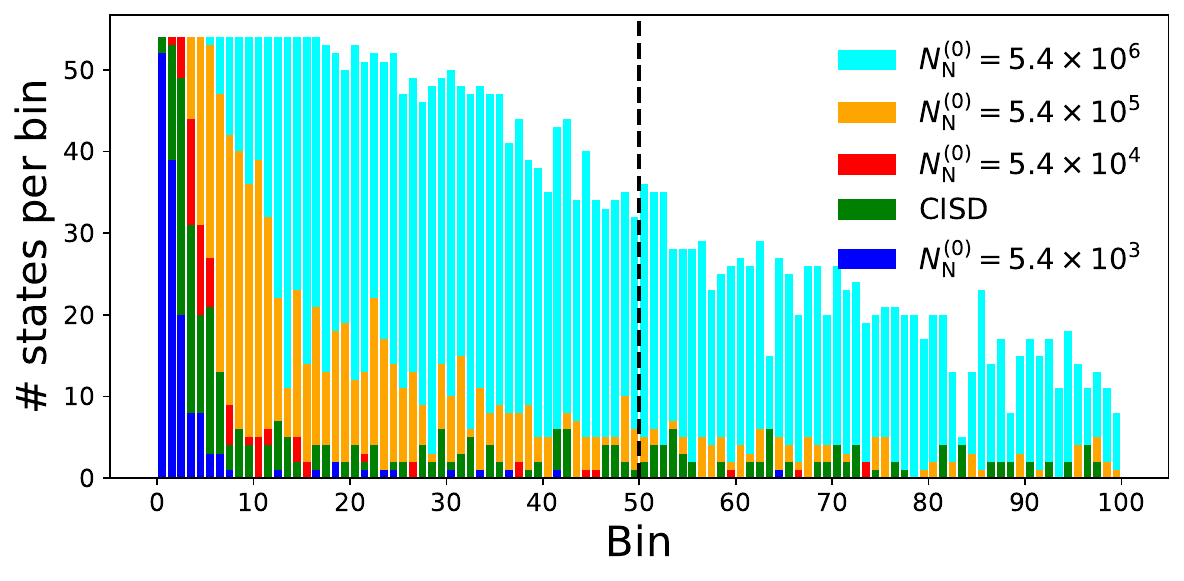}}  
    \hspace{2em}
    \subfigure[\ $\text{C}_2\text{H}_4$ final iteration]{\includegraphics[width=0.45\linewidth]{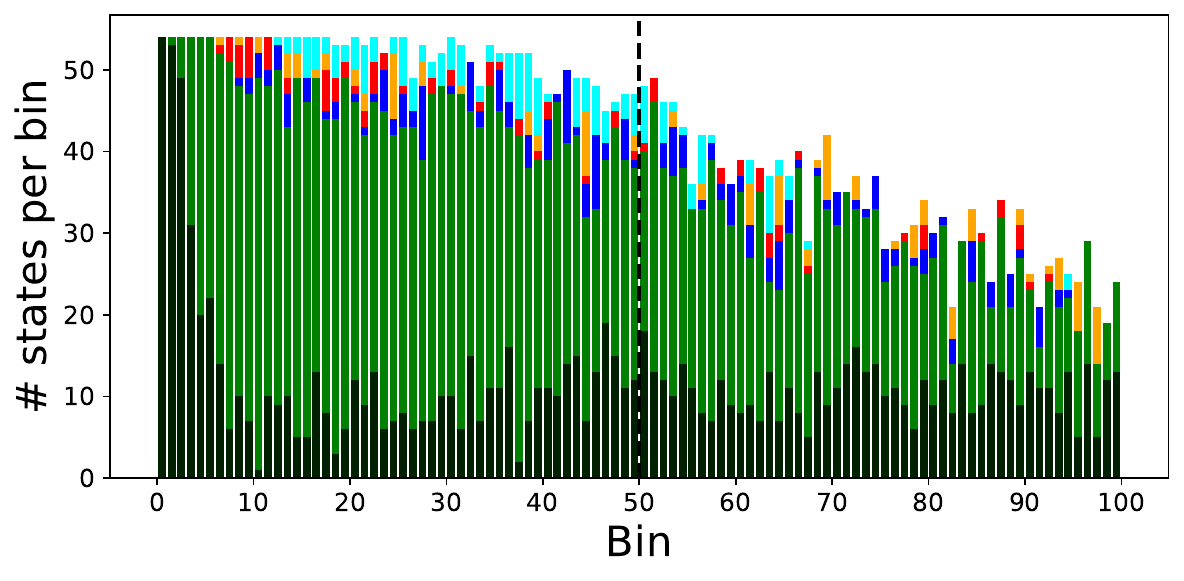}}
    \caption{Visualization of, for each curve in Figs. \ref{fig:convergence_vs_cheat_C2H2_10_6} and \ref{fig:convergence_vs_cheat_C2H4_10_6} minus the HF and raw CISD curve, how well they fill up the $2N_\text{CA}$ most highly contributing configurations to the $\text{C}_2\text{H}_2$ and $\text{C}_2\text{H}_4$ ground states. The configurations are organized into $100$ bins of $\frac{2N_\text{CA}}{100}$ states each, with the most dominant states appearing on the left-hand-side. The vertical, dashed line separates the $N_\text{CA}$ most important states from the next $N_\text{CA}$. [(a) and (c)]: the filling from the original GS approximation (CISD or $N_\text{N}^{(0)}$ GS samples). [(b) and (d)]: the filling by the end of the procedure (the energy stops decreasing within the bounds of our setup). The CISD results employ the technique of temperature scaling. The black bars in (b) and (d) indicate states obtained after only one iteration using the HF state, with temperature scaling $\beta = 0.1$.}
    \label{fig:C2H2_important_states_ts_w_hf}
\end{figure}

The filling of the highest-contributing configurations is visualized in Fig. \ref{fig:C2H2_important_states_ts_w_hf} (a) and (b). In (a), we notice that while many dominating states are already picked out at iteration zero, getting close to obtaining the first $N_\text{CA}=800$ states (which alone generate a subspace ground-state energy within chemical accuracy) is only possible when taking very many samples of the exact ground state. In (b), however, we can see that our procedure does an excellent job at getting these first $800$ states filled in, with differences between initial setups being hard to distinguish. Indeed, the Figure suggests no real advantage in starting with $N_\text{N}^{(0)}=1.4\times 10^4$, over using $N_\text{N}^{(0)}=1.4\times 10^3$ or even the CISD state, with temperature scaling. This suggests that ``obvious" correlations already built in to the network architecture before training, i.e. the inductive bias \cite{Mitchell80,valleperez2019}, gives a boost to these weaker approximations whereas these correlations might be lost with intermediate sample sizes. To see how this could be the case, consider Fig. \ref{fig:C2H2_important_states_ts_w_hf} (b). Here, instead of plotting the filling-of-configurations for the Hartree--Fock state, we instead show the filling achieved after only one iteration to illustrate how it can result in decent performance after several iterations. By using temperature scaling with $\beta=0.1$, some physically relevant configurations, initially assigned very low probabilities by the neural network, become visible in the sampling. Meanwhile, the surprisingly poor performance of $N_\text{N}^{(0)}=1.4\times 10^4$ may be attributed to the large collective weight carried by unimportant states, which leads to wasteful sampling and prevents important but less-dominant configurations from appearing in the data.

% Some testing has shown that just starting with the Hartree--Fock state, when using temperature scaling, leads to similar or sometimes even better performance than when starting with CISD at iteration zero.

\begin{figure}[t]
    \centering
    \includegraphics[width=0.95\linewidth]{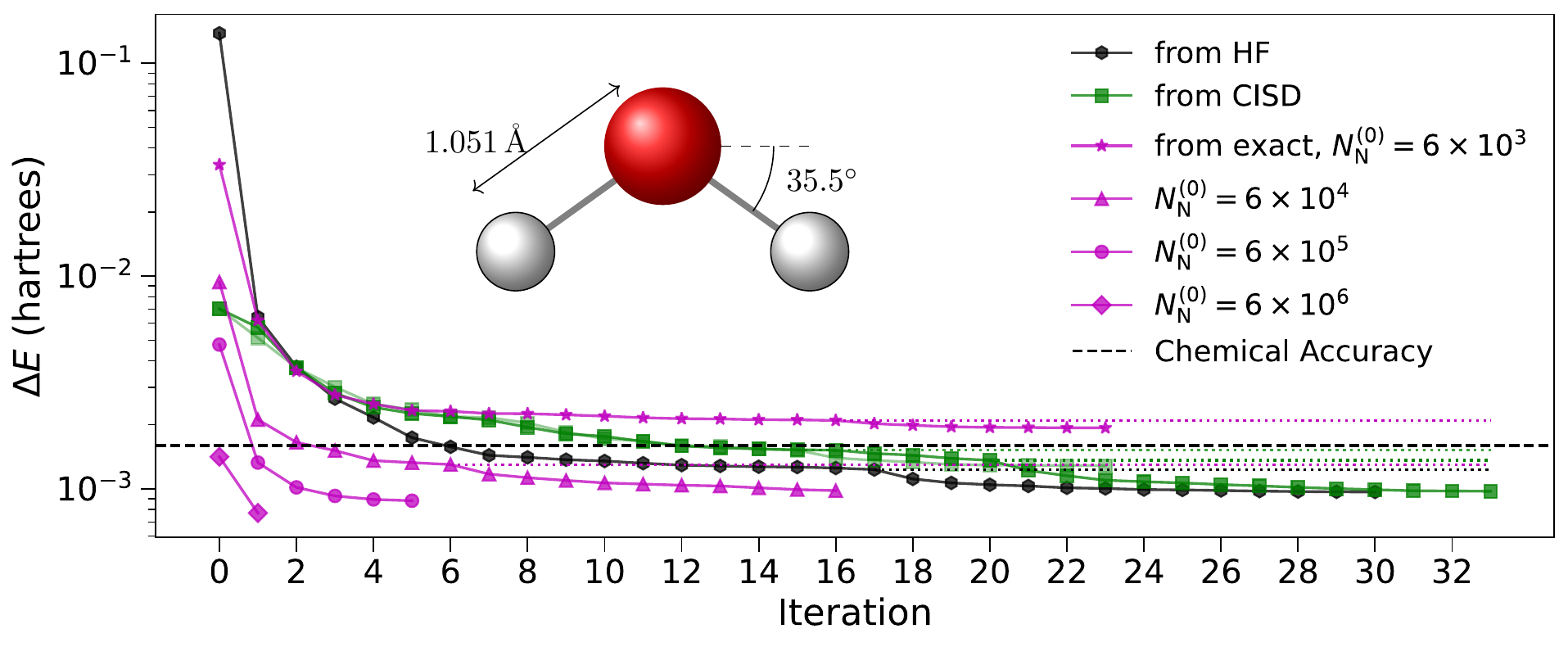}
    \caption{Ground state energy error $\Delta E$ \textit{vs} iteration number obtained for the $\text{H}_2\text{O}$ molecule using the 6-31g basis, with displayed molecular geometry the same as that from \cite{Choo2020}. The solid/faded, green curve corresponds to the raw/temperature-scaled CISD-initiated curve. When sampling from the exact ground state, we consider four cases where the number of shots differ by an order of magnitude. We take $N_\text{N}=6\times 10^6$ and $N_\text{U}=4000$ in all curves. The horizontal dashed lines in all but the bottom two curves indicate a switch in the network architecture, where the number of layers and features are doubled, and $N_\text{T}$ increases from $10^4$ to $10^5$. Only the larger model is used in the bottom two curves. Additional hyper-parameters: two masked-dense layers, four features per bit, and dropout rate 0.05 in the smaller model, each of which are doubled in the larger model. ADAM learning rate 0.001. The temperature-scaled CISD curve takes $\beta=0.6$ for the first iteration. HF takes $\beta=0.1,0.6$ for the first two iterations.}
    \label{fig:H2O}
\end{figure}

Next, we look at the $\text{H}_2\text{O}$ molecule in the 6-31g basis, which has 26 molecular spin-orbitals (available single-electron wavefunctions) and $414,441$ symmetry-satisfying configurations. Here we have $N_\text{CA}=2,000$ and $N_\text{N} = 6\times 10^6$. As before we set $N_\text{U}=2N_\text{CA}=4,000$. The results are depicted in Fig. \ref{fig:H2O}, and show that while the CISD-based results are capable of outperforming GS-based results for very small sample sizes, e.g. $N_\text{N}^{(0)}=6\times 10^3$, it still struggles even to match the energy achieved by $N_\text{N}^{(0)}=6\times 10^4$, taking far longer to reach that energy. As before, the HF curve performs even better than that for CISD, but it also struggles to show improvement over $N_\text{N}^{(0)}=6\times 10^4$ and in this case requires many iterations to do so.

\begin{figure}[t]
    \includegraphics[width=0.95\linewidth]{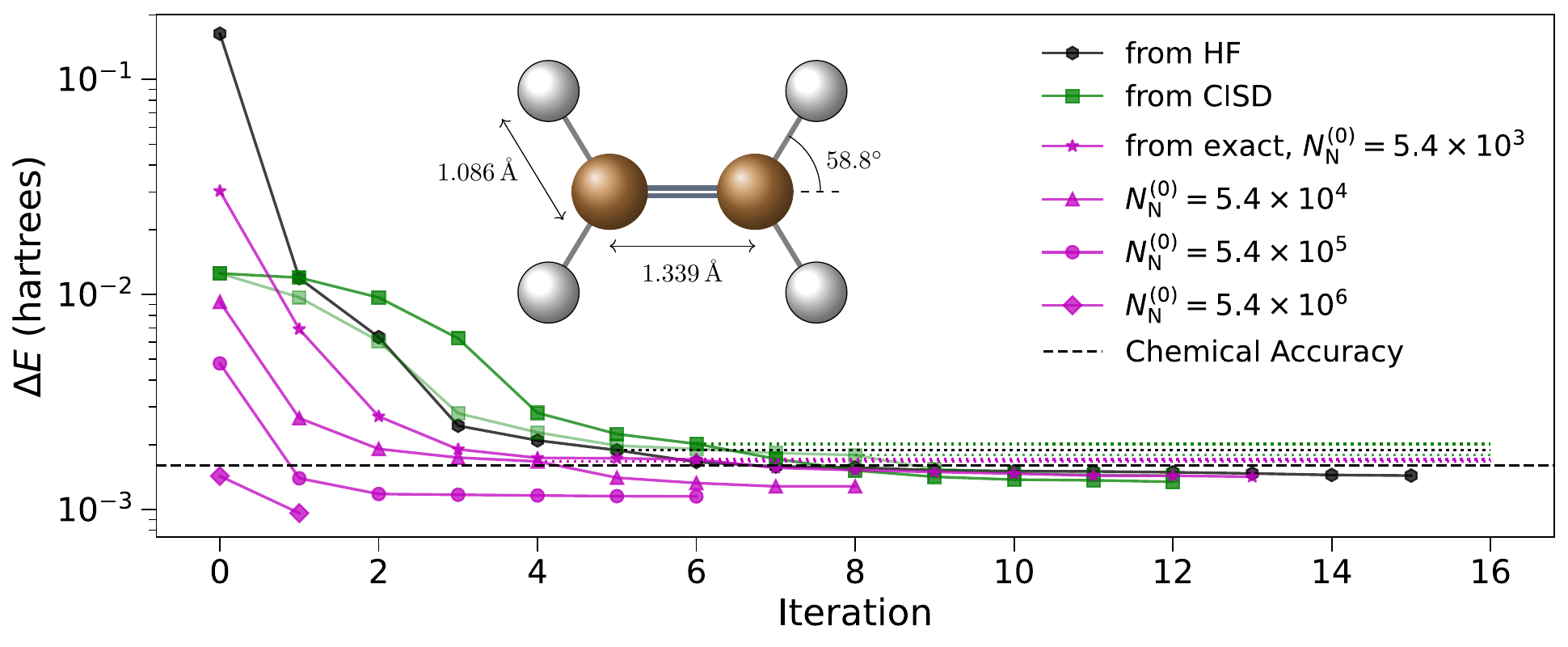}
    \caption{Ground state energy error $\Delta E$ \textit{vs} iteration number obtained for the $\text{C}_2\text{H}_4$ molecule in the STO-3G basis set, with displayed molecular geometry the same as the first geometry considered in \cite{Machado1997}. The solid/faded, green curve corresponds to the raw/temperature-scaled CISD-initiated curve. When sampling from the exact ground state, we consider four cases where the number of shots differ by an order of magnitude. We take $N_\text{N}=5.4\times 10^6$ and $N_\text{U}=5400$ in all curves. The horizontal dashed lines in all but the bottom two curves indicate a switch in the network architecture, where the number of layers and features are doubled, and $N_\text{T}$ increases from $10^4$ to $10^5$. Only the larger model is used in the bottom two curves. Additional hyper-parameters: two masked-dense layers, four features per bit, and dropout rate 0.05 in the smaller model, each of which are doubled in the larger model. ADAM learning rate 0.001. The temperature-scaled CISD curve takes $\beta=0.4$ for the first iteration. HF takes $\beta=0.1,0.6$ for the first two iterations.}
    % , with $N_N=10^6$ and $N_U=5,000$. The two magenta ``Exact State" plots take $N_T=10^4$ and $N_T=10^6$, respectively. The temperature scaling $\beta_0=0.4$ is applied to the initial state after the first iteration, while the ARNN temperature scaling $\beta$ is varied in early iterations to encourage rapid energy minimization. Horizontal, dotted lines originate at final points where no sampled states are forcibly retained (see text for details). Additional hyper-parameters: four masked-dense layers, eight features per bit. Dropout rate 0.1 and ADAM learning rate 0.001. }
    \label{fig:convergence_vs_cheat_C2H4_10_6}
\end{figure}

\begin{figure}[t]
    \includegraphics[width=0.95\linewidth]{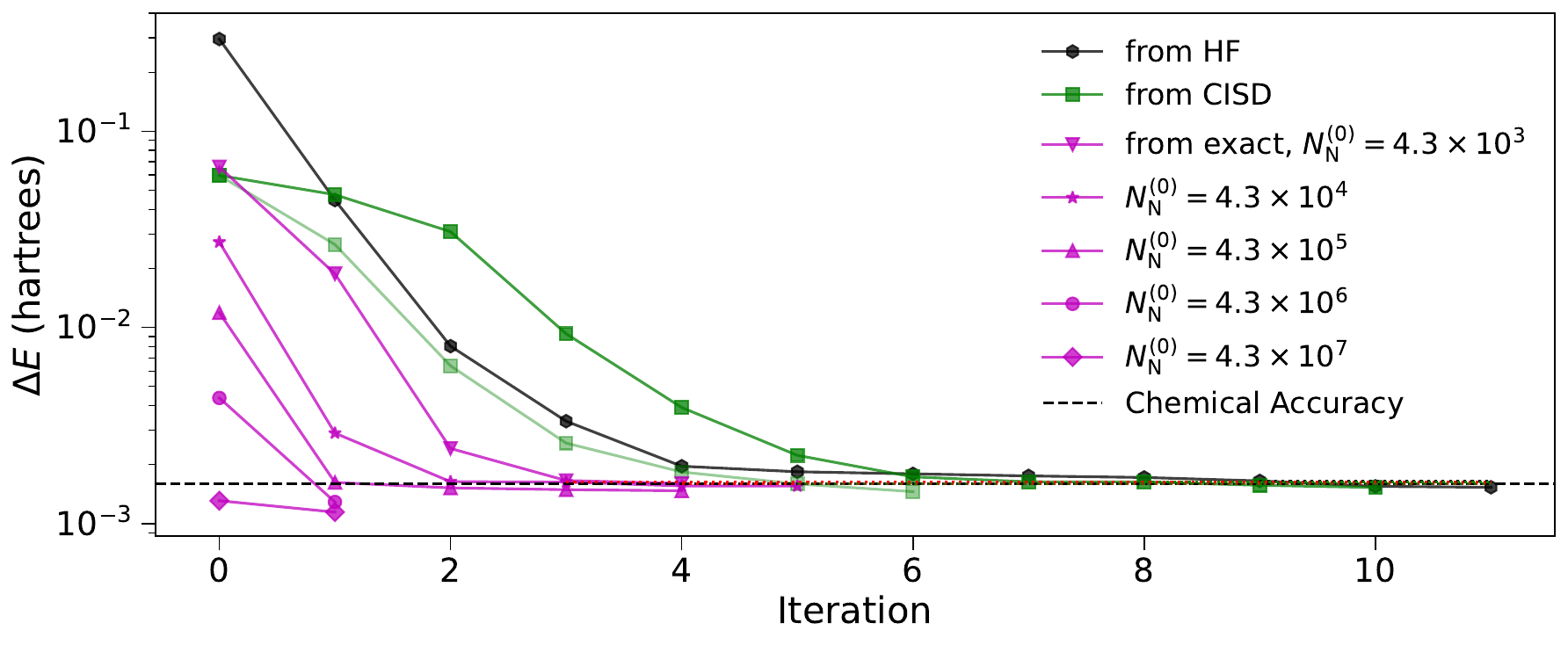}
    \caption{Ground state energy error $\Delta E$ \textit{vs} iteration number obtained for the $\text{C}_2$ molecule, with a bond length of $1.271$ \AA \cite{Johnson2002,Herzog2023}. The solid/faded, green curve corresponds to the raw/temperature-scaled CISD-initiated curve. When sampling from the exact ground state, we consider five cases where the number of shots differ by an order of magnitude. We take $N_\text{N}=4.3\times 10^7$ and $N_\text{U}=34,000$ in all curves. The horizontal dashed lines in all but the bottom three curves indicate a switch in the network architecture, where the number of layers and features are doubled, and $N_\text{T}$ increases from $10^4$ to $10^5$. Only the larger model is used in the bottom three curves. Additional hyper-parameters: two masked-dense layers, four features per bit, and dropout rate 0.05 in the smaller model, each of which are doubled in the larger model. ADAM learning rate 0.001. The temperature-scaled CISD curve takes $\beta=0.8$ for the first iteration. HF takes $\beta=0.15$ for the first iteration.}
    \label{fig:C2}
\end{figure}

We see similar sets of curves for the $\text{C}_2\text{H}_4$ and $\text{C}_2$ molecules, Figs. \ref{fig:convergence_vs_cheat_C2H4_10_6} and \ref{fig:C2}. We take $\text{C}_2\text{H}_4$ in the STO-3G basis, which gives 28 spin-orbitals and $1,131,361$ symmetry-respecting configurations. It has $N_\text{CA}=2,700$ and $N_\text{N}=5.4\times 10^6$. For a visualization of the filling of the $2N_\text{CA}$ most-contributing configurations, see Fig. \ref{fig:C2H2_important_states_ts_w_hf} (c) and (d). $\text{C}_2$ is not stable at room temperature \cite{Miyamoto2020}, yet it remains a good benchmark molecule \cite{Choo2020,Herzog2023} that allows us to study a system with $36$ spin-orbitals in the 6-31g basis set. It has $43,114,512$ symmetry-respecting configurations. Results for $\text{C}_2$, with $N_\text{CA}=17,000$ and $N_\text{N}=4.3\times 10^7$, are given in Fig. \ref{fig:C2}.

% \begin{table}[t]
%     \centering
%     \begin{tabular}{|c|c|c|c|}
% \hline
% Molecule & Symmetry Respecting & Number Preserving & Total \\
% \hline
% $\text{C}_2\text{H}_2$ & $17.2$ &  $2.23$ & $0.083$\\
% $\text{H}_2\text{O}$ & $14.5$ &  $3.62$ & $0.0894$\\
% $\text{C}_2\text{H}_4$ & $4.77$ &  $0.599$ & $0.0201$\\
% $\text{C}_2$ & $0.997$ &  $0.125$ & $6.26\times 10^{-4}$\\
% \hline
% \end{tabular}
% \caption{Caption}
% \label{fig:num_states_table}
% \end{table}

Table \ref{tab:num_states_table} tracks the ratios of the values $N_\text{U}$, which thus far we've set to $2N_\text{CA}$, to the number of configurations in the Hilbert space with three levels of symmetry constraints. In all three cases, we see that this ratio decreases rather quickly, supporting the argument that for many molecules the wavefunction has some sparsity in the computational basis, at least in the sense of achieving a desired accuracy. While the column labeled ``Symmetry Respecting" tracks what this ratio is within the correct symmetry sectors (electron number and point-group symmetry), we must still recall that the neural network itself acts on the entire Fock space (all fermionic occupations) without any symmetries applied. The final column therefore indicates that the neural network is able to successfully identify important configurations within the correct symmetry sector while at the same time still sampling configurations with incorrect quantum numbers (the proportion of states we need to discard does indeed decrease as we approach convergence, thus much of this extra noise is fortunately reduced). These ratios are expected to improve further if the orbitals, fixed in our case according to the Hartree--Fock calculation, are themselves optimized as is done in CASSCF \cite{McArdle2020}.

% \renewcommand{\arraystretch}{1.5}
% \setlength{\tabcolsep}{8pt}
% \begin{table}[t]
%     \centering
%     \begin{tabular}{|c|c|c|c|c|}
% \cline{3-5}
% \multicolumn{2}{c|}{} & \multicolumn{3}{c|}{\textbf{$\mathbf{N_U/\text{\# states}}$}} \\    
% \hline
% Molecule/Basis & \# Qubits & Symmetry Respecting & Number Preserving & Total \\
% \hline
% $\text{C}_2\text{H}_2$/STO-3G & 24 & $2.03\times 10^{-2}$ &  $2.55\times 10^{-3}$ & $9.54\times 10^{-5}$\\
% $\text{H}_2\text{O}$/6-31g & 26 & $9.65\times 10^{-3}$ &  $2.41\times 10^{-3}$ & $5.96\times10^{-5}$\\
% $\text{C}_2\text{H}_4$/STO-3G & 28 & $4.77\times 10^{-3}$ &  $5.99\times 10^{-4}$ & $2.01\times 10^{-5}$\\
% $\text{C}_2$/6-31g & 36 & $7.89\times 10^{-4}$ &  $9.87\times 10^{-5}$ & $4.95\times 10^{-7}$\\
% \hline
% \end{tabular}
% \caption{Values of the ratios $\frac{N_\text{U}}{\text{\# states}}$ for each molecule considered where, numbering from left to write, the \# of states is (1) the total number of determinants with the correct electron number and point-group representation, (2) the total number of determinants with the correct electron number only, and (3) the total number of possible Fock states in the full Hilbert space.}
% \label{tab:num_states_table}
% \end{table}

\renewcommand{\arraystretch}{1.5}
\setlength{\tabcolsep}{6pt}
\begin{table}[t]
    \centering
    \begin{tabular}{|c|c|c|c|c|c|c|c|}
\cline{3-8}
\multicolumn{2}{c|}{} & \multicolumn{6}{c|}{\textbf{$\mathbf{\text{\# of states}}$}\quad\big\vert\quad\textbf{$\mathbf{\frac{N_U}{\text{\# of states}}}$}} \\    
\hline
\multicolumn{1}{|c|}{Molecule/Basis} & \multicolumn{1}{c|}{\# Qubits} & \multicolumn{2}{c|}{Symmetry Respecting}& \multicolumn{2}{c|}{Number Preserving}& \multicolumn{2}{c|}{Total} \\    
\hline
% Molecule/Basis & \# Qubits & Symmetry Respecting & Number Preserving & Total \\
% \hline
$\text{C}_2\text{H}_2$/STO-3G & 24 & $78,992$ & $2.03\times 10^{-2}$ & $627,264$ &  $2.55\times 10^{-3}$ & $16,777,216$ & $9.54\times 10^{-5}$\\
$\text{H}_2\text{O}$/6-31g & 26 & $414,441$ & $9.65\times 10^{-3}$ & $1,656,369$ &  $2.41\times 10^{-3}$ & $67,108,864$ & $5.96\times10^{-5}$\\
$\text{C}_2\text{H}_4$/STO-3G & 28 & $1,131,361$ & $4.77\times 10^{-3}$ & $9,018,009$ &  $5.99\times 10^{-4}$ & $268,435,456$ & $2.01\times 10^{-5}$\\
$\text{C}_2$/6-31g & 36 & $43,114,512$ & $7.89\times 10^{-4}$ & $344,622,096$ &  $9.87\times 10^{-5}$ & $68,719,476,736$ & $4.95\times 10^{-7}$\\
\hline
\end{tabular}
\caption{Values for the \# of states and the ratios $\frac{N_\text{U}}{\text{\# of states}}$ for each molecule considered where, numbering from left to right, the \# of states is (1) the total number of configurations with the correct electron number and point-group representation, (2) the total number of configurations with the correct electron number only, and (3) the total number of possible configurations in the full Fock space.}
\label{tab:num_states_table}
\end{table}

\begin{figure}[t]
    \centering
    \includegraphics[width=0.95\linewidth]{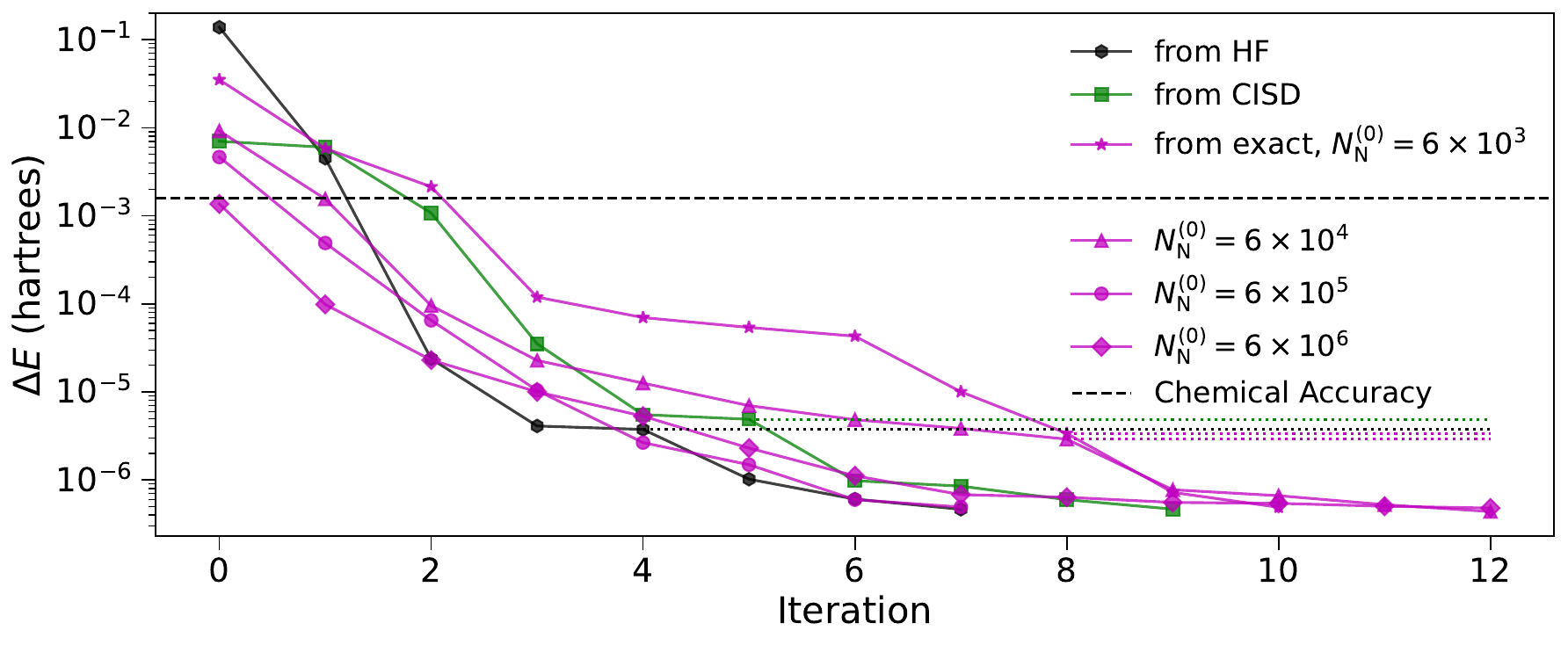}
    \caption{Ground state energy error $\Delta E$ \textit{vs} iteration number obtained for the $\text{H}_2\text{O}$ molecule using the 6-31g basis. When sampling from the exact ground state, we consider four cases where the number of shots differ by an order of magnitude. We take $N_\text{N}=6\times 10^6$, but do not set a value for $N_\text{U}$, i.e. the size of the sampled subspace is set by the total number of unique configurations identified. The horizontal dashed lines in all but the bottom two curves indicate a switch in the network architecture, where the number of layers and features are doubled, and $N_\text{T}$ increases from $10^4$ to $10^5$. Only the larger model is used in the bottom two curves. Additional hyper-parameters: two masked-dense layers, four features per bit, and dropout rate 0.05 in the smaller model, each of which are doubled in the larger model. ADAM learning rate 0.001. For CISD, we observe only the raw case, i.e. that without any temperature scaling. HF takes temperature scaling $\beta=0.1$ for the first iteration.}
    \label{fig:H2O_keep_all_states}
\end{figure}

It turns out that having the exact state on hand (or a good guiding state containing enough information about correlation effects) is only expected to be useful in more restrictive settings, namely when the size $N_\text{U}$ of the sampled subspace is to remain under control. To see this, consider again the water molecule in Fig. \ref{fig:H2O}. This time we do not set a value $N_\text{U}$ and allow the size of the sampled subspace to be controlled only by the total number of unique configurations sampled. We obtain Fig. \ref{fig:H2O_keep_all_states}.

In this case we do not see too much of an advantage in sampling from the exact ground state, as starting with the HF or CISD approximation gives excellent convergence to accurate energies in just a handful of iterations. On the other hand, this Figure gives us some insight into the true power of this procedure, and it is not hard to imagine that for larger system sizes the computational cost becomes too high if no cap $N_\text{U}$ is enforced. Based on our results above, enforcing this cap would most likely remove any advantage of starting from the HF or CISD approximation, which favors the utilization of a sufficiently-sampled guiding state.

\begin{figure}[t]
    \includegraphics[width=0.95\linewidth]{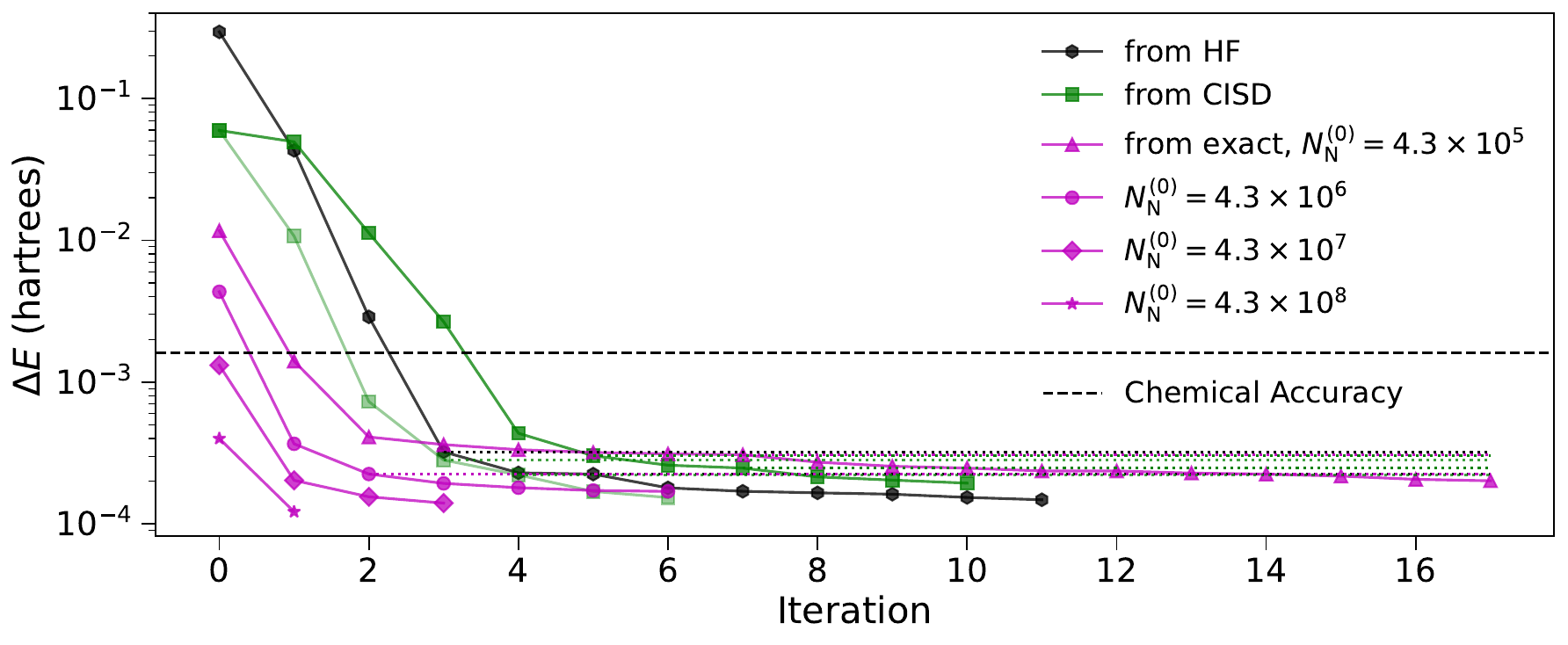}
    % \subfigure[]{\includegraphics[width=0.5\linewidth]{figs/C2_200000_more_samples.pdf}}
    \caption{Same procedure as Fig. \ref{fig:C2}, except that we set $N_\text{U}=200,000$, and that we allow a second enhancement in the network architecture (with model switches indicated by horizontal, dotted lines as well). This final structure contains six masked-dense layers, twelve features per bit, dropout rate 0.15, and $N_\text{T}=10^6$ during training. Unlike Fig. \ref{fig:C2}, we test GS sample sizes ranging from $N_\text{N}^{(0)}=4.3\times 10^5$ to $N_\text{N}^{(0)}=4.3\times 10^8$. The temperature-scaled CISD curve takes $\beta=0.6$ for the first iteration. HF takes $\beta=0.15,0.8$ for the first and second iterations. }
    % For the number of network samples, we have (a) $N_\text{N}=4.3\times 10^7$ and (b) $N_\text{N}=28\times 10^9$ (the latter lowered by a factor of $10$ in some iterations due to limited memory).  Temperature-scaled curves are not included in (b) due to memory constraints.}
    \label{fig:C2_200000}
\end{figure}

To check whether this suggestion has any grounds, we turn back once more to the $\text{C}_2$ molecule in the 6-31g basis. Removing the cap $N_\text{U}$ in this case is indeed very expensive, resulting in massive sampled subspaces. Instead of restricting all the way to $N_\text{U}=2N_\text{CA} = 34,0000$ as before, we take $N_\text{U} = 200,000$, see Fig. \ref{fig:C2_200000}. To allow accuracy to improve in accordance with the larger subspace, we admit an additional switch in the size of the NN model after the larger network from the previous examples ceases to lower the energy. As expected, we find that the HF- and CISD-based curves perform better against the GS-based curves, when compared with Fig. \ref{fig:C2}. However, in both cases their converged energies are still unable to match those of the curves that took a very large number of samples from the ground state, namely the curves corresponding to $N_\text{N}^{(0)}=4.3\times 10^7$ and $N_\text{N}^{(0)}=4.3\times 10^8$. Therefore, as one might have suspected, while a guiding state is advantageous for smaller sampled subspaces, an attempt to increase accuracy by increasing the size of the sampled subspace still, in turn, demands more samples $N_\text{N}^{(0)}$ from the original guiding state.

% Note that in Figure \ref{fig:C2_200000} (b), we increase the sample size closer to the expected number needed to see the $200,000$ ``best" states in the original, exact GS.

\section{Conclusions} \label{section:Conclusion}
Accurately determining ground-state energies and wavefunctions remains a central challenge in quantum chemistry, especially for strongly-correlated systems and large molecular structures. In this work, we have introduced an algorithm that uses auto-regressive neural networks to guide subspace expansion for Selected Configuration Interaction (SCI). This Neural Quantum State (NQS) approach does not require variational Monte Carlo (VMC) and therefore has the potential to sidestep its convergence challenges. When trained on limited data from approximate ground states, even restricted to just computational-basis measurements, these models generalize effectively, identifying previously unseen electronic configurations that meaningfully contribute to the true ground state. Upon performing exact diagonalization in the subspace of identified configurations, this results, after just a few iterations, in energies that surpass chemical accuracy across several benchmark molecular systems. Temperature scaling further enhances the approach by both smoothing out data from previous iterations for improved training, as well as helping to capture many important configurations during sampling in early iterations, speeding up convergence toward the desired accuracy. The iterative capability of our procedure allows a degree of flexibility in the nature of the initial data, giving us some freedom in choosing the input approximation for the ground state. We have seen how this, in turn, complicates the question of when a good guiding state becomes advantageous.  

Heading toward the strongly-correlated regime, then, we wish to confirm that a well-designed and properly-implemented guiding state holds a clear advantage over the simple approximations that have thus far proved effective. Additionally, improvements and modifications to the algorithm itself may be required in this regime. It is reasonable to assume that MCSCF techniques could assist not only in constructing a reference active-space wavefunction but also in identifying more useful orbital bases for second quantization. Our approach thus far has been implemented entirely classically, yet in the spirit of exploring strongly-correlated systems it leaves open the possibility for augmentation by quantum algorithms. As computational basis measurements on their own may not provide sufficient training data in this regime, it will be interesting to see to what extent improvements are seen as we further tailor our algorithm to effectively accommodate measurement data associated with rotated bases \cite{Bennewitz2022,Huang_2020,Torlai2020,Gunlycke2024}.

Looking ahead, additional future work will focus on scaling the method to larger molecules and richer basis sets, with attention to how computational costs scale with system size. But perhaps one of the most important reasons for the continued investigation of this method is its potential for analyzing samples from a noisy quantum simulator, allowing it to act as an effective error-mitigation \cite{Torlai_bases,Bennewitz2022,gunlycke2025} and variance-reduction technique. If data from our initial approximation contains significant noise but reflects enough information about the underlying physics for the neural-network approach to notice and encode it, then our procedure may be able to keep refining this noisy data up until chemical accuracy is achieved. This possible edge in the post-processing of quantum data, along with the other considerations we have just discussed, will be key in establishing our ARNN-based approach as a robust, hybrid-ready framework for quantum chemistry.

\section*{Acknowledgements}
This work has been supported by the Office of Naval Research through the U.S. Naval Research Laboratory. This work was supported in part by a grant of computer time from the Department of Defense (DoD) High Performance Computing Modernization Program (HPCMP). The computations were performed on the Raider system at the Air Force Research Laboratory DoD Supercomputing Resource Center (AFRL DSRC). S.T. thanks the National Research Council Research Associateship Programs for support during his post-doctoral tenure at NRL.
\bibliographystyle{unsrt}
\bibliography{main}
\appendix
\section{Neural Quantum States and Variational Monte Carlo}\label{appendix:VMC}
\begin{figure}[t]
    \centering
    \includegraphics[width=0.35\linewidth]{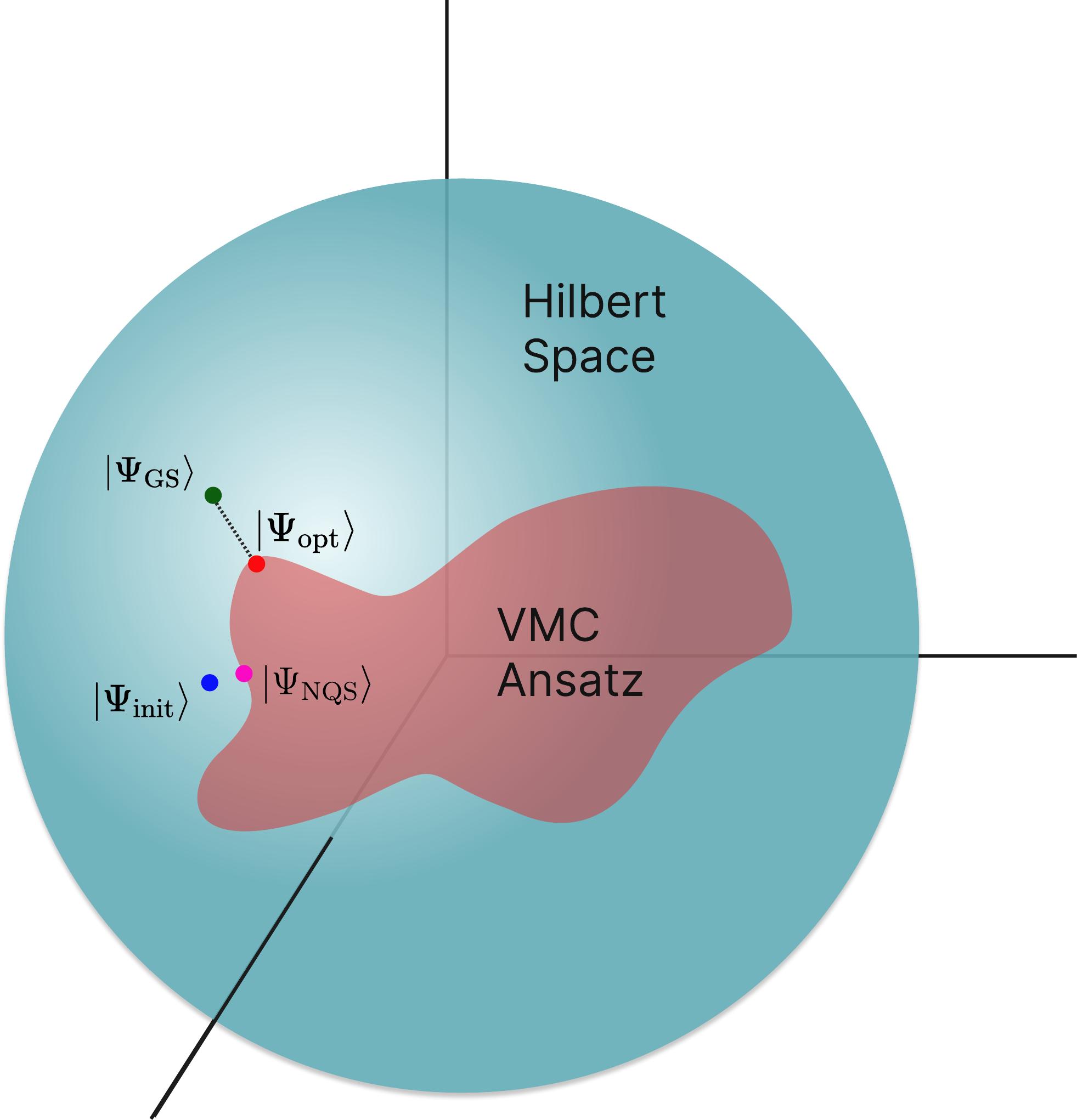}
    \caption{Visual representation of the VMC method. An NQS is trained within the variational manifold formed by the VMC ansatz, using the energy functional $\langle H\rangle$. The result of the optimization, $\ket{\Psi_\text{opt}}$ approximates the true ground state, $\ket{\Psi_{\text{GS}}}$. Note that as in Ref. \cite{Bennewitz2022}, training can be improved by initializing the network parameters using data from some $\ket{\Psi_\text{init}}$. The initialized NQS is indicated by the point labeled $\ket{\Psi_\text{NQS}}$.}
    \label{fig:VMC_diagram}
\end{figure}

In this Appendix we review the Neural Quantum State (NQS) approach to Variational Monte Carlo (VMC). The energy functional (cost function) we seek to minimize is
\begin{equation}\label{eq:cost_fn}\frac{\langle \Psi\vert H\vert\Psi\rangle}{\langle\Psi\vert\Psi\rangle}=\frac{1}{\sum_{n} \vert \Psi\left(n\right)\vert^2}\sum_{n,n'}\Psi^*\left(n\right)H_{nn'}\Psi\left(n'\right)\end{equation}
where $\Psi(n) = \langle n\vert\Psi\rangle$ is not assumed normalized over configurations $n$ and $H_{nn'}=\langle n\vert H\vert n'\rangle$. The $\ket{n}$ form a computational basis, which in the main text we took to be fermionic Fock-state configurations corresponding to Slater determinants over single-particle molecular wavefunctions (molecular spin-orbitals) obtained from the Hartree--Fock procedure. The minimization is performed with respect to a set of variational parameters $\alpha_k$ which define an NQS ansatz for $\ket{\Psi}$ within the Hilbert space, see Fig. \ref{fig:VMC_diagram}.

We can rewrite Eq. \eqref{eq:cost_fn} as
\be \label{eq:CLT_VMC} \frac{\langle \Psi\vert H\vert\Psi\rangle}{\langle\Psi\vert\Psi\rangle} = \sum_{n}\left[\sum_{n'}H_{nn'}\frac{\Psi\left(n'\right)}{\Psi\left(n\right)}\right]\left[\frac{\vert\Psi(n)\vert^2}{\sum_{n''} \vert\Psi(n'')\vert^2}\right]=\sum_{n}E_\text{loc}\left(n\right)P\left(n\right)\equiv \mathbb E_P\left(E_\text{loc}\right),\ee
where $P(n)$ is the Born probability  of configuration $n$ (in the computational basis) given by 
\be \label{eq:prob} P\left(n\right)\equiv \frac{\vert\Psi(n)\vert^2}{\sum_{n'} \vert\Psi(n')\vert^2}\ee
and
\be \label{eq:local_estimator} E_\text{loc}\left(n\right)\equiv \sum_{n'}H_{nn'}\frac{\Psi\left(n'\right)}{\Psi\left(n\right)}\ee
is known as the local energy estimator \cite{Vicentini2022}. Assuming $H$ is a sparse matrix in the $\ket{n}$ representation, the value of the function $E_\text{loc}(n)$ can be computed efficiently given $n$. The accuracy of Eq. \eqref{eq:CLT_VMC} relies on the Central Limit Theorem (CLT) and thus the variance of $E_\text{loc}$ over the Hilbert space. 
% Unless $\ket{\Psi}$ is an exact eigenstate, this variance can be large given the presence of $\Psi$ in the denominator of \eqref{eq:local_estimator} \cite{}.

Minimization of \eqref{eq:cost_fn} requires us to compute its derivative. It is straightforward to show that
\be\label{eq:cost_grad} \frac{\partial}{\partial \alpha_k} \frac{\langle \Psi\vert H\vert\Psi\rangle}{\langle\Psi\vert\Psi\rangle}=2\Re\{\mathbb{E}_P\left(\mathcal{O}_k^* E_\text{loc}\right)-\mathbb{E}_P\left(\mathcal{O}_k^*\right)\mathbb{E}_P\left(E_\text{loc}\right)\}\ee
where 
\be \mathcal{O}_k(n)=\frac{1}{\Psi(n)}\frac{\partial \Psi(n)}{\partial \alpha_k} = \frac{\partial \log\Psi(n)}{\partial \alpha_k},\ee
which can be computed by applying automatic differentiation to the network. Eq. \eqref{eq:cost_grad} may be used directly in SGD \cite{Saito2017}, although common practice applies Stochastic Reconfiguration \cite{Carleo2017,Vicentini2022} and solves the equation
\be\label{eq:sr} \overset{\leftrightarrow}{S}\delta\vec\alpha = \vec F\ ,\ \ \ee
where $\vec \alpha$ is the vector of network parameters, $\vec F$ is the force vector $\mathbb{E}_P(\mathcal{\vec O}^* E_\text{loc})-\mathbb{E}_P(\mathcal{\vec O}^*)\mathbb{E}_P(E_\text{loc})$ and
\be S_{k,k'} = \mathbb{E}_P(\mathcal{O}_k^* \mathcal{O}_{k'})-\mathbb{E}_P(\mathcal{O}_k^*)\mathbb{E}_P(\mathcal{O}_{k'})\ee
is the quantum geometric tensor (QGT) describing the geometry of the variational manifold. Eq. \ref{eq:sr} is derived from imaginary-time evolution by finding the parameter changes necessary to minimize the Fubini-Study distance between the quantum state, at each small time step, and the NQS.

In any case, Eqs. \eqref{eq:CLT_VMC}, \eqref{eq:cost_grad}, and \eqref{eq:sr} contain averages with respect to the probability distribution $P(n)$. Clearly, proper convergence of VMC hinges on the variance properties of $E_\text{loc}$ and $\mathcal{O}$. The distribution we sample from can be generated via Markov-chain Monte Carlo (MCMC). Namely, we move randomly between different configurations (up to what the Hamiltonian allows direct transitions between) and according to the Metropolis--Hastings algorithm \cite{Metropolis1953,Hastings1970} decide to accept or reject the move between $\ket{n}$ and $\ket{n'}$ based on the probability ratio
\be\label{eq:prob_ratio} \Bigg\vert\frac{\Psi\left(n'\right)}{\Psi\left(n\right)}\Bigg\vert^2.\ee
$\Psi\left(n\right)$ is the un-normalized wave-function for configuration $n$. Since we take ratios, we need not worry about the normalization $\sum_{n} \vert\Psi(n)\vert^2$. Notice also that
\eqref{eq:prob} is an actual probability distribution (real and non-negative), and therefore VMC is not plagued by a sign problem in the usual sense. We benefit further in the case where the network is auto-regressive, as we can then perform direct, unbiased sampling of the distribution without worrying about auto-correlation effects \cite{MuellerKrumbhaar1973,Wu2021}.

\section{Method Improvements}\label{appendix:method_improvements}
\subsection{Previous Simplifications}
Our algorithm makes use of exact diagonalization in a sampled subspace, based on the assumption of sparsity of the Born probabilities in the computational basis \cite{Ivanic2001,Anderson2018}. To see how the situation can break down for the more general case, we borrow the expression for the expectation value $\bra{\Psi}H\ket{\Psi}$ that is used in VMC, expressed in a form resembling what is constructed in Ref. \cite{Gunlycke2024}. We let 
\be \ket{\Psi(\vec\theta)} = \frac{e^{i\lambda\left(\vec\theta\right)}}{\sqrt{\Lambda(\vec\theta)}}\ket{\Psi_0}\ee 
where $\lambda(\vec\theta)$ is an operator diagonal in the computational basis and dependent on a set of variational parameters $\vec\theta$, and $\Lambda(\vec\theta)$ enforces normalization. In the reference, $\ket{\Psi_0}$ is a state measured on a quantum computer. To relate to our case, we should think of it as some neural-network approximation of the ground state that can be sampled efficiently. The VMC expectation may be written as $\langle H\rangle=\frac{\Upsilon(\vec\theta)}{\Lambda(\vec\theta)}$, where
\begin{align}
    \Upsilon(\vec\theta) &= \sum_{n}\left[\sum_{n'}H_{nn'}e^{-i\lambda^*(n;\vec\theta)}e^{i\lambda(n';\vec\theta)}\frac{\Psi_0(n')}{\Psi_0(n)}\right]\vert\Psi_0(n)\vert^2\ ,\nonumber\\
\Lambda(\vec\theta) &= \sum_{n}e^{-2\text{Im}\lambda(n;\vec\theta)}\vert\Psi_0(n)\vert^2\end{align}
We estimate these quantities with statistical averages over a sample distribution $\mathcal{S}$ in the computational basis:
\begin{align}\label{eq:upslam_statave}
    \Upsilon(\vec\theta) &\approx \frac{1}{\vert\mathcal{S}\vert}\sum_{n\in \mathcal{S}}\left[\sum_{n'}H_{nn'}e^{-i\lambda^*(n;\vec\theta)}e^{i\lambda(n';\vec\theta)}\frac{\Psi_0(n')}{\Psi_0(n)}\right]N_{n}\ ,\nonumber\\
\Lambda(\vec\theta) &\approx \frac{1}{\vert\mathcal{S}\vert}\sum_{n\in \mathcal{S}}e^{-2\text{Im}\lambda(n;\vec\theta)}N_{n}\end{align}
where $N_{n}$ is the number of times configuration $n$ is sampled. This form is essentially an effort to sample some reference distribution described by $\Psi_0$ and perform a re-weighting procedure such that the ansatz $\Psi$ can be optimized by varying $\lambda$. This is similar to what is done in Refs. \cite{Umrigar1988,Foulkes2001}.
% This can only be made equivalent to exact diagonalization in the sampled subspace $\mathcal{S}$ if we assume
% that we can safely truncate the sum in brackets to just the $n'\in\mathcal{S}$, which might follow from assuming $\vert\Psi_0{(n')}\vert\ll\vert\Psi_0(n)\vert$ for $n'\notin\mathcal{S}$ which just sets us up to miss important correlation effects and essentially signals that a compact description of the wavefunction is sufficient to reach chemical accuracy.

Let us now suppose that an approximate description of the wavefunction over the sampled subspace $\mathcal{S}$ is sufficient to reach chemical accuracy, so that we can force $\ket{\Psi}$ (via $\lambda(n;\vec\theta)\to i\infty$ for $n\notin\mathcal{S}$) to have support only over this subspace:
\be \ket{\Psi} \overset{!}{\approx} \mathcal{N}_\mathcal{S}(\vec\theta)\sum_{n\in\mathcal{S}}e^{i\lambda\left(n;\vec\theta\right)}\Psi_0(n)\ket{n} \ee
Then in this variational subspace we truncate the sum in brackets to just the $n'\in\mathcal{S}$, and rewrite Eq. \eqref{eq:upslam_statave} as
\begin{align}\label{eq:upslam_var}
    \Upsilon(\vec\theta) &\approx \frac{1}{\vert\mathcal{S}\vert}\sum_{n,n'\in\mathcal{S}}\left(e^{-i\lambda^*(n;\vec\theta)}\Psi_0^*(n)\right)H_{nn'}\left(e^{i\lambda(n';\vec\theta)}\Psi_0(n')\right)\frac{N_{n}}{\vert\Psi_0(n)\vert^2}\ ,\nonumber\\
\Lambda(\vec\theta) &\approx \frac{1}{\vert\mathcal{S}\vert}\sum_{n\in\mathcal{S}}\vert e^{i\lambda(n;\vec\theta)}\Psi_0(n)\vert^2\frac{N_{n}}{\vert\Psi_0(n)\vert^2}\end{align}

% Our second assumption is
% \begin{center} (2) \textit{The state $\ket{\Psi}$ need only be realized in the sampled subspace $\mathcal{S}$}\end{center}
% This assumption was crucial in justifying diagonalization in only the sampled subspace, but such an assumption has no foundation in very strongly correlated systems where $\ket{\Psi}$ has broad support. 
We use the general form of $\lambda$ to take
\be \Psi_0(n)\to e^{i\tilde\lambda(n;\vec\theta)}\sqrt{\frac{N_{n}}{\vert\mathcal{S}\vert}}\ee
and absorb $\tilde\lambda$ into $\lambda$. We find
\begin{align} \label{eq:ED_equiv}
    \Upsilon(\vec\theta) &\approx \sum_{n,n'\in\mathcal{S}}\left(e^{-i\lambda^*(n;\vec\theta)}\sqrt{\frac{N_{n}}{\vert\mathcal{S}\vert}}\right)H_{nn'}\left(e^{i\lambda(n';\vec\theta)}\sqrt{\frac{N_{n'}}{\vert\mathcal{S}\vert}}\right)\ ,\nonumber\\
\Lambda(\vec\theta) &\approx \sum_{n\in\mathcal{S}}\Bigg\vert e^{i\lambda(n;\vec\theta)}\sqrt{\frac{N_{n}}{\vert\mathcal{S}\vert}}\Bigg\vert^2=1\end{align}
Minimizing $\langle H\rangle$ then becomes equivalent to exact diagonalization in the sampled subspace, with all amplitudes and phases fixed by $\lambda(\vec\theta)$. This method also gives us a way to throw out all phase information associated with $\ket{\Psi_0}$. For wavefunctions with compact enough support for subspace diagonalization, this is not an issue, as the recovery of phases through $\lambda$ becomes an efficient process. This is why our neural network did not need to be complex-valued. 

While the energy minimum of Eq. \eqref{eq:ED_equiv} is achieved with a state $\ket{\Psi}$ having support only in a compact subspace $\mathcal{S}$, in very strongly correlated systems it may need to have very broad support. Unlike our approach thus far, Eq. \eqref{eq:upslam_statave} does not explicitly prevent this from happening (VMC does not assume sparsity). A proper extension of our method would likely relax the truncation assumption and perform a variational procedure over expressions \eqref{eq:upslam_statave} with respect to $\lambda(\theta)$, instead of performing diagonalization. That is, taking $\ket{\Psi_i}$ to be the reference at iteration $i$, we would compute
\be\label{eq:E_i_def} E_i = \underset{\vec\theta}{\text{min}}\frac{\mathbb{E}_{P_i}\Big[h(n;\vec\theta)\Big]}{\mathbb{E}_{P_i}\Big[l(n;\vec\theta)\Big]}\ ,\ \ \ee
where $P_i(n) = \vert \Psi_i(n)\vert^2$ and 
\begin{gather} h(n;\vec\theta) = \sum_{n'}H_{nn'}e^{-i\lambda^*(n;\vec\theta)}e^{i\lambda(n';\vec\theta)}\frac{\Psi_i(n')}{\Psi_i(n)}\ ,\ \ \nonumber\\l(n;\vec\theta) = e^{-2\text{Im}\lambda(n;\vec\theta)}\label{eq:h_and_l_def}\end{gather}
are the functions whose averages over $P_i(n)$ are estimated. Noting that it would be natural to simply absorb the exponential in $\lambda$ into $\Psi_i(n)$, the latter of which we have represented with an ARNN, it may be possible that the only natural extension of our method is Variational Monte Carlo, in some form. Even so, future work might reveal that it is useful to let $\lambda$ be some separate neural network whose parameters $\vec\theta$ we optimize with respect to sampling data from the ARNN.

\subsection{Complex-valued Networks}
% Extensions of our approach would most likely require sampling in additional measurement bases in order to appropriately capture phase information, yet thus far we have mostly only discussed what additional measures may be required for iteration zero. More specifically, while the first of Eqs. \eqref{eq:h_and_l_def} would require a direct amplitude calculation for $\Psi_0$, one can compute the energy expectation in other ways, given experimental data (see Ref. \cite{Gunlycke2024}). However, the main theme of this work was to train neural networks in order to generalize the measurement data, and with an ARNN representation of $\ket{\Psi_q}$ we may still use Eqs. \eqref{eq:E_i_def} and \ref{eq:h_and_l_def} directly as we go beyond iteration zero.

Generalizations of our approach would most likely require sampling in additional measurement bases and non-sparse (e.g. sampled subspace) variational procedures to appropriately capture phase information. It follows that we should expect to design an ARNN that is complex-valued. This can be realized by a straightforward generalization of the ARNN shown in Fig. \ref{fig:our_ARNN}. Recall that the second-to-last layer represents the normalized conditional wavefunction $\Psi_q$ for each bit $q$, and therefore each bit has two ``features" whose values, or more precisely, squared exponentials, sum to one. If we add an additional real-valued feature to each bit, and then apply a linear layer over these extra values, we can obtain a single, real scalar which represents the phase of the bitstring's wavefunction value \cite{Bennewitz2022}. Fig. \ref{fig:our_ARNN} would need an additional bit to encode the known configuration of the final physical bit, but this would not affect the conditional probabilities of the physical bits (nor the real parts of their conditional log amplitudes).

The components of an NQS in a non-computational basis $\mathcal{B}$ are given by \cite{Bennewitz2022} 
\be\label{eq:non_computational_amplitude} \Psi\left(n|\mathcal{B}\right) = \sum_{n'}{}_{\mathcal{B}}\langle n|n'\rangle\Psi\left(n'\right) \ee
with $\ket{n}_\mathcal{B}$ a configuration in basis $\mathcal{B}$. Therefore, the network must be trained in a way to account for measurement results obtained for each $\mathcal{B}$. In determining which $\mathcal{B}$ we need as well as what proportion of the data corresponds to each of them, reasonable approaches include randomly choosing measurement bases \cite{Torlai_bases}, weighting them by relevance to the Hamiltonian \cite{Tilly2022}, or even applying importance sampling using knowledge obtained from classical shadows \cite{kokaew2024}. Every time a basis $\mathcal{B}$ and configuration $n$ are chosen, one applies Eq. \ref{eq:non_computational_amplitude} to evaluate the network (or its gradient). In the case of electronic-structure Hamiltonians equipped with measurements restricted to single-qubit rotations, this equation contains at most $16$ terms.

% Having already identified a good modification operator $\lambda(\theta)$ after a full iteration, we require a technique to sample the resulting ground-state approximation ($e^{i\lambda}$ applied to $\ket{\Psi_q}$, where $\ket{\Psi_q}$ could be an ARNN) so that we may proceed to the next iteration. Unfortunately, the lack of a sparse representation in the computational basis, or even an auto-regressive NN representation which produces unbiased samples in every basis $\mathcal{B}$, likely limits \TODO{[are we sure?]} sampling capabilities to MCMC methods. As mentioned in Appendix \ref{appendix:VMC}, this does not guarantee an efficient simulation of unbiased sampling of the the probability distributions derived from the Born rule \cite{MuellerKrumbhaar1973,Wu2021}. Therefore, it may be necessary to investigate additional alternatives in future work. On top of this, one must be able to identify advantages distinguishing this approach, whatever form it takes, and VMC with ``correlated sampling" \cite{Foulkes2001}. The latter does not make use of the KL divergence \eqref{eq:KL_div}.

\end{document}